\documentclass[11pt,a4paper]{article}
\usepackage[utf8x]{inputenc}
\usepackage[T1]{fontenc}
\usepackage{mathptmx} 

\usepackage[margin=25truemm]{geometry}
\usepackage[pdftex]{graphicx} 
\usepackage{calc} 
\usepackage{authblk}
\usepackage{bm}
\usepackage{enumitem} 
\usepackage{braket,mathtools,amsmath,amssymb,subcaption}
\usepackage{authblk}
\usepackage[all]{nowidow} 
\usepackage{url}
\newcommand{\cc}[2]{({#1},{#2})}
\usepackage{lipsum} 

\usepackage{framed} 
\usepackage[framed]{ntheorem}
\newframedtheorem{frm-thm}{Theorem}
\usepackage{hyperref}
\hypersetup{colorlinks,linkcolor=blue}

\def\m{\mathcal}

\def\s{\sigma}

\def\b{\boldsymbol}
\def\g{\eta}
\def\Ao{A_0}
\def\Bo{B_0}
\def\Ai{A_I}
\def\Bi{B_I}
\begin{document} 
\title{Entanglement entropy of higher rank topological phases}

\author{Hiromi Ebisu}
\affil{Department of Physics and Astronomy,
Rutgers University, Piscataway, NJ 08854,
USA}
\maketitle
\begin{abstract}
We study entanglement entropy of unusual $\mathbb{Z}_N$ topological stabilizer
codes which admit fractional excitations with restricted mobility constraint in a manner akin to fracton topological phases.
It is widely known that the sub-leading term of the entanglement entropy of a disk geometry in conventional topologically ordered phases is related to the total number of the quantum dimension of the fractional excitations.
We show that, in our model, such a relation does not hold, i.e, 
the total number of the quantum dimension varies depending on the system size, whereas the sub-leading term of the entanglement entropy takes a constant number irrespective to the system size. 
We give a physical interpretation of this result in the simplest case of the model. More thorough analysis on the entanglement entropy of the model on generic lattices is also presented.
\end{abstract}

\section{Introduction}
Since their discovery, topologically ordered phases have had a tremendous impact on the
condensed matter physics community~\cite{Tsui,Laughlin1983,Kalmeyer1987,wen1989chiral}. These phases have drawn a lot of attention mainly because of the
exotic fractionalized excitations they carry, called anyons~\cite{Laughlin1983,leinaas1977theory}. For practical purposes, they may find application in quantum computers~\cite{dennis2002topological,KITAEV20032,Mong}.\par

While a lot of efforts have been made to understand these phases from both of theoretical and experimental point of view, recently new types of topologically ordered phases have been introduced, referred to as the fracton topological phases~\cite{chamon,Haah2011,PhysRevB.92.235136}. These phases 
cannot be described by the preexisting frameworks of the topologically ordered phases, due to the distinctive feature that the phases are sensitive to the local geometry of the system. Key insight to intuitively understand these phases is a mobility constraint is imposed on the fractional excitations, giving rise to the sub-extensive ground state degeneracy~(GSD).
Establishing a complete theoretical frameworks of these phases is an active research topic. 
\par 
One of the attempts to handle this problem is to introduce new types of symmetries, \textit{multipole symmetries}. 
The multipole symmetry, in particular, the $U(1)$ multipole symmetry, is the generalization of the global $U(1)$ symmetry; a theory is invariant under the global phase rotation depending on the spatial coordinate in polynomial form. 
As an example, in the case of a scalar theory which respects the global and dipole $U(1)$ symmetries, the Lagrangian is invariant under $\Phi\to e^{ia+ibx}\Phi$, where $a$ and $b$ represent constants, and $x$ does the spatial coordinate~\cite{Pretko:2018jbi}. Investigating topological and fracton topological phases in this view point of the new symmetry has been recently started~\cite{PhysRevB.66.054526,Seiberg:2019vrp,griffin2015scalar,Pretko:2018jbi,PhysRevX.9.031035,Jain:2021ibh,PhysRevB.106.045112}. \par

Spurred by this new interest in the interplay between topological phases and multipole symmetries, in~\cite{ebisu2209anisotropic}, a simple model of unconventional topological phases with dipole symmetry (which is one of the multipole symmetries) was studied. 
New feature of the model is that it admits dipole of the Wilson loops of the fractional excitations, giving rise to the unusual ground state degeneracy (GSD) dependence on the system size. In this work, we dub topological phases with multipole symmetries \textit{higher rank topological phases}.
While the GSD has been thoroughly investigated in the model in~\cite{ebisu2209anisotropic} as well as other higher rank topological phases, other physical properties of the models have yet to be explored. To address this problem, 
in this paper, we investigate the interplay between the higher rank topological phases and the \textit{entanglement entropy}, which is an important physical quantity to diagnose the quantum entanglement.\par
A plethora of progress has been made in understanding quantum entanglement of conventional topologically ordered phases. It is well known that the entanglement entropy of the 2D topologically ordered phases in the disk bipartite geometry $A$, $S_A$ scales as~\cite{KitaevPreskill2006TEE,LevinWen2006TEE}
\begin{equation}
    S_A\sim\alpha l-\gamma+\cdots.\label{ee}
\end{equation}
    The first term, which is the leading term, is the so-called area-law term, proportional to the perimeter of the disk geometry $l$ with $\alpha$ being non-universal constant, whereas the second constant term, which is the sub-leading term, is of particularly importance as it is universal number,
    referred to as the \textit{topological entanglement entropy}. It is rewritten as
    \begin{equation}
       \gamma=\log\biggl(\sqrt{\sum_ad^2_a}\biggr),\label{ee2}
    \end{equation}
    where $d_a$ represents the quantum dimension of the anyon labeled by $a$. [$\cdots$ in~\eqref{ee} represents terms that vanish in the limit of $l\to\infty$.]

We study entanglement entropy of 
various geometries of the subsystems in the unconventional topological phase with the dipole symmetry. 
By making use of the formalism of the entanglement entropy for stabilizer codes~\cite{Hamma2005}, jointly with the one in the graph theory, we show that the entanglement entropy is described by the same form as~\eqref{ee}, yet Eq.~\eqref{ee2}, which is an important relation between entanglement entropy and the total quantum dimension of fractional excitations, does \textit{not} hold. Indeed, while the total quantum dimension varies depending on the geometry of the lattice, the topological entanglement entropy~$\gamma$ takes the constant number, irrespective to the lattice.
We give an intuitive interpretation of the result by focusing on the simplest case by setting $N=2$. 
We further study the entanglement entropy of the higher rank topological phase on the generic lattices.
We show that the entanglement entropy of the sub-graph consists of the two types of the terms, the leading order area law term and the sub-leading constant term, both of which depend on the the number of vertices surrounding the sub-graph and the invariant factors of the Laplacian, which is the matrix describing how the vertices are connected in the graph. The result is summarized in~Table~\ref{table1} and Theorem~\ref{th}.\par
There has been an intimate relation between quantum entanglement and graph theory. For instance, multiparty entangle states is described by a graph and its properties are studied based on graph theory, see e.g.,~\cite{PhysRevA.69.062311,PhysRevLett.86.910}. 
To the best of our knowledge, this paper is the first work to address the interplay between graph theory and quantum entanglement in the context of (new type) topological stabilizer models by studying the Laplacian. Our work would comply with recent interest in unconventional topological phases, in particular, higher rank topological phases with multipole symmetries, from view point of the quantum entanglement and graph theory. 
\par
The rest of the paper is organized as follows. In Sec.~\ref{sec2}, we review the model and its properties studied in~\cite{ebisu2209anisotropic}. After reviewing the model, in Sec.~\ref{sec3}, we calculate entanglement entropy of various geometries of subsystems. 
In Sec.~\ref{sec4}, we give an intuitive understanding of our result on the entanglement entropy.
In Sec.~\ref{sec7}, we further investigate the entanglement entropy of a sub-graph in the case where the model in defined on a generic lattice.
 We make a brief comment on other cases of topological phases in Sec.~\ref{sec5}. 
Finally, we conclude our work in Sec.~\ref{sec6}. Technical details are relegated to appendices.


\section{Stabilizer model}\label{sec2}
In this section, we review the model constructed by stabilizers given in~\cite{ebisu2209anisotropic}. Also, we go over physical properties of the model. As we mentioned in the previous section, the model exhibits unusual behavior of the excitations compared with the conventional topologically ordered phases, yielding the unusual GSD dependence on the lattice.
\subsection{Hamiltonian}
To start, we introduce 2D square lattice where we 
place two types of $N$-qubit state ($\mathbb{Z}_N$ clock states) on each vertex and vertical link. 
 The first clock states are located at vertices of the lattice whereas the second ones are at vertical links. 
We label the coordinate of the first clock states by~$(x,y)$ and the ones of the second clock states are denoted by $(x,y+\frac{1}{2})$, where the second element corresponds to the links between vertices located at~$(x,y)$ and $(x,y+1)$.
We term these two types of the clock states clock state with type 1 and clock state with type 2.
\par
\begin{figure}[h]
    \begin{center}
        
       \includegraphics[width=0.39\textwidth]{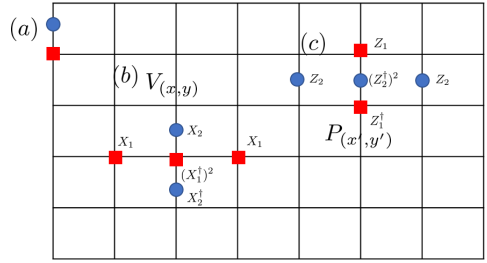}
\end{center}
       \caption{ (a)Two types of $\mathbb{Z}_N$ clock states, defined on each vertex (red square) and vertical link (blue dot).  
      (b)(c) Two types of terms introduced in~\eqref{pp}.
 }\label{fig1}

   \end{figure}

We represent basis of the two types of the clock states as $\ket{\omega}_1$ and $\ket{\omega}_2$ with $\omega$ being $N$-th root of unity, i.e, $\omega=e^{2\pi i/N}$ [red square and blue dot in Fig.~\ref{fig1}(a)], and $\mathbb{Z}_N$ shift and clock operators (they become Pauli operators when $N=2$) of the first and second clock states as $\{Z_i,X_i\}\;(i=1,2)$. Here we have introduced the subscript $i=1,2$ to distinguish operators that act on the clock clock states with type 1 and the ones with type 2.
These operators satisfy
the following relation ($I_i$ denotes the identity operator)
\begin{equation}
  X_i^N=Z_i^N=I_i,\;  Z_i\ket{\omega}_i=\omega\ket{\omega}_i, \; X_iZ_j=\omega Z_jX_i\delta_{i,j}. \label{relation}
\end{equation}

With these notations, 
we define following two types of operators at each vertex and link
\begin{eqnarray}
V_{(x,y)}\vcentcolon=X_{2,(x,y+1/2)}X_{2,(x,y-1/2)}^{\dagger}(X_{1,(x,y)}^{\dagger})^{2}X_{1,(x+1,y)}X_{1,(x-1,y)},\nonumber\\
P_{(x,y+1/2)}\vcentcolon=Z_{1,(x,y+1)}Z_{1,(x,y)}^{\dagger}(Z_{2,(x,y+1/2)}^{\dagger})^{2}Z_{2,(x-1,y+1/2)}Z_{2,(x+1,y+1/2)}.\label{pp}
\end{eqnarray}
These terms are portrayed in Fig.~\ref{fig1}(b)(c). We interchangeably call these two terms, $V_{(x,y)}$ and~$P_{(x,y+1/2)}$, \textit{vertex} and \textit{plaquette operators}, respectively. 
It is straightforward to check that each term in~\eqref{pp} commute, forming the stabilizer group.
 Hamiltonian is defined by
\begin{equation}
    H=-\sum_{(x,y)}V_{(x,y)}-\sum_{(x,y+1/2)}P_{(x,y+1/2)}+h.c.\label{hamiltonian}
\end{equation}
This model shares the same feature as the $\mathbb{Z}_N$ toric code~\cite{KITAEV20032}. The ground state is the stabilized state: the ground state $\ket{\Omega}$ satisfies
\begin{equation}
    V_{(x,y)}\ket{\Omega}=P_{(x,y+1/2)}\ket{\Omega}=\ket{\Omega}\nonumber \;\;\forall\;V_{(x,y)}, P_{(x,y+1/2)}. 
\end{equation}
Also, our model admits two types of fractional excitations when the condition $V_{(x,y)}=1$ or $P_{(x,y+1/2)}=1$ is violated at a vertex or link. The crucial difference between our model and the toric code is that 
that each terms in the Hamiltonian involves not only nearest neighboring sites but also next nearest neighboring ones. As a consequence, the fractional excitations are subject to mobility constraint, giving rise to unusual GSD dependence on the system size.
We dub the topological phases with this feature \textit{higher rank topological phases}. Throughout this paper, we consider the model placed on the torus geometry with length in the $x(y)$ direction being $n_x(n_y)$.
\subsection{Simplest case: $N=2$ }\label{sec2b}
\begin{figure}[h]
    \begin{center}
      \begin{subfigure}[h]{0.24\textwidth}
  \includegraphics[width=\textwidth]{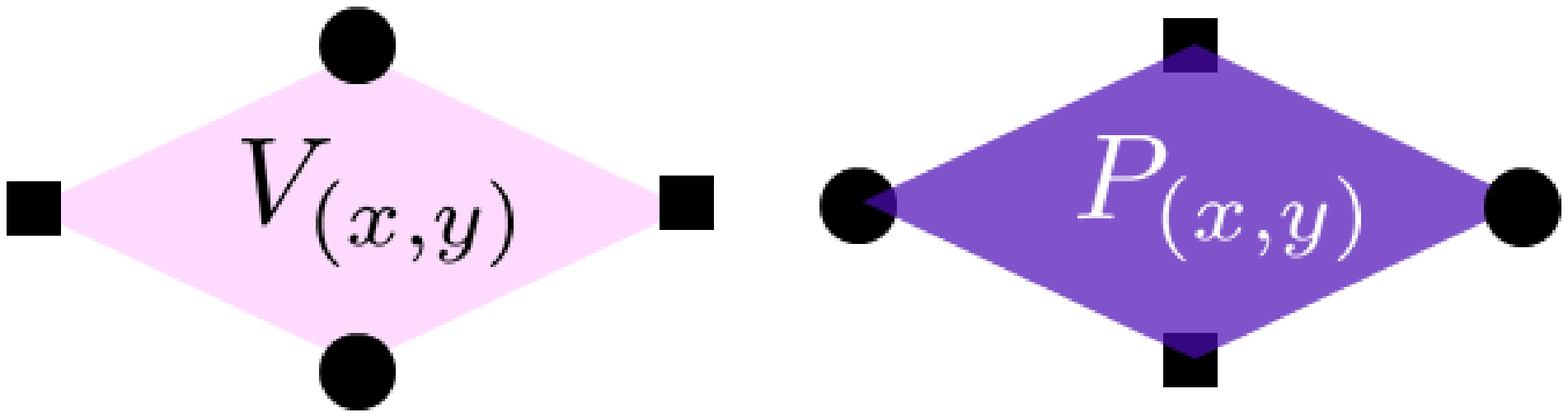}
         \caption{}\label{sfa}
             \end{subfigure}
        \begin{subfigure}[h]{0.66\textwidth}
    \includegraphics[width=\textwidth]{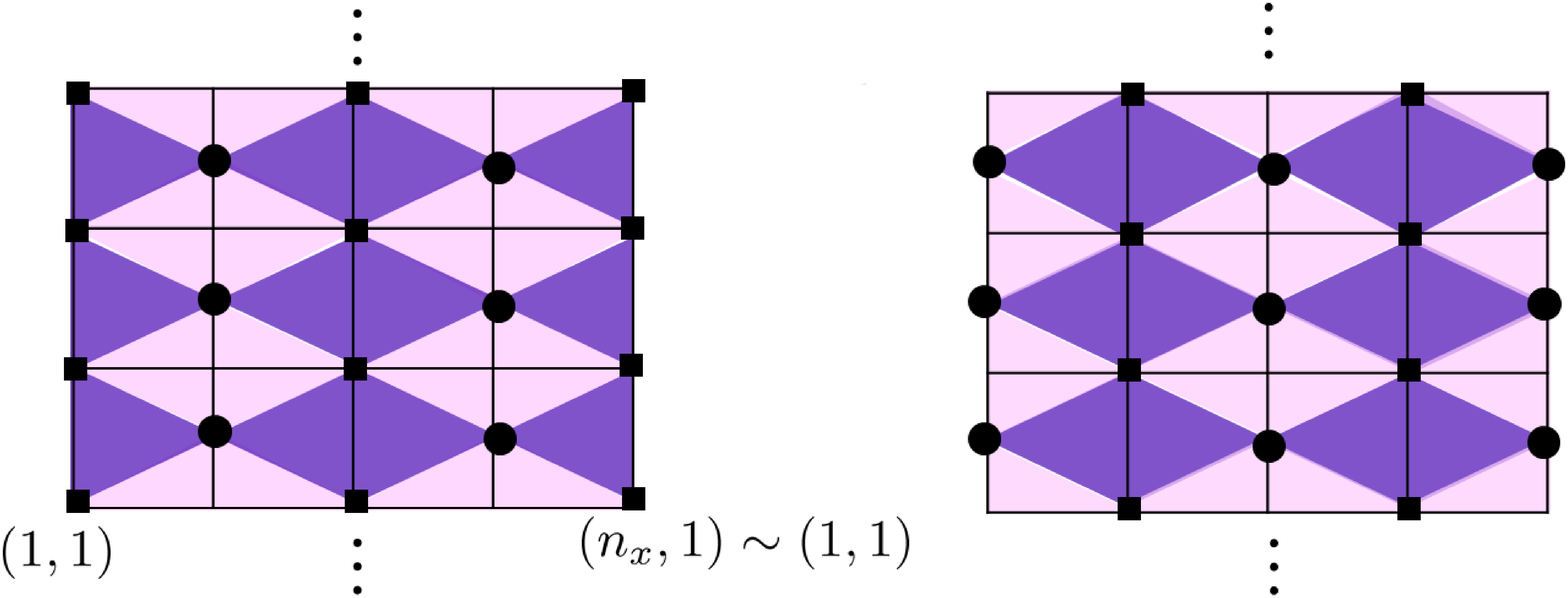}
         \caption{}\label{sfb}
             \end{subfigure} 
\end{center}
\caption{(a)Two terms defined in~\eqref{ppa} constituting the Hamiltonian in the case of $N=2$. (b)~Configurations of stabilizers corresponding to~\eqref{group} in the case of $n_x$ even. In this case, the model is decomposed into two $\mathbb{Z}_2$ toric codes. The periodic boundary condition is imposed in the $x$-direction so that left and right edge are identified. 
The left [right] geometry corresponds to configuration of the vertex and plaquette terms which belong to (i)[(ii)] defined in~\eqref{group}.
Such decomposition is not possible in the case of $n_x$ being odd. We set the coordinate of the vertex at the bottom left to be $(1,1)$.  }
   \end{figure}

Before disusing the model for generic case of $N$, 
we consider the simplest case by setting $N=2$
to make more intuitive understanding of our model~\eqref{hamiltonian}. The argument presented here will be useful to interpret our result on the entanglement entropy~(Sec.~\ref{sec4}). In this case, the terms defined in~\eqref{pp} become
\begin{eqnarray}
   V_{(x,y)}&=&X_{2,(x,y+1/2)}X_{2,(x,y-1/2)}X_{1,(x-1,y)}X_{1,(x+1,y)},\nonumber\\
P_{(x,y+1/2)}&=&Z_{1,(x,y+1)}Z_{1,(x,y)}Z_{2,(x-1,y+1/2)}Z_{2,(x+1,y+1/2)}.\label{ppa}
\end{eqnarray}
One can regard each term as the plaquette term in a rhombus shape, as portrayed in Fig.~\ref{sfa}.
The Hamiltonian~\eqref{hamiltonian} with~\eqref{ppa} resembles the $\mathbb{Z}_2$ toric code. However there is a crucial deference between the two: each term~\eqref{ppa} involves \textit{next} nearest neighboring Pauli operators in the $x$-direction, as opposed to the regular toric code where each term consists of nearest neighboring Pauli operators. 
Due to this property, one can separate the mutually commuting terms~\eqref{ppa} into two groups without considering the boundary:
\begin{equation}
    \text{(i)}\{V_{(2m,y)},P_{(2m^\prime-1,y^\prime+1/2)}\},\;\; \text{(ii)}\{V_{(2m-1,y)},P_{(2m^\prime,y^\prime+1/2)}\}\;\;(m,m^\prime\in\mathbb{Z}).\label{group}
\end{equation}
  Hence, without thinking the boundary, the model that we describe with $N=2$ amounts to be two decoupled $\mathbb{Z}_2$ toric codes.\footnote{It is proved that for prime $p$, all translation invariant $\mathbb{Z}_p$ stabilizer codes in 2D are equivalent to copies of toric codes by a local Clifford circuit of constant depth ~\cite{haah2021classification}. It would be interesting to see whether one can decompose our model into copies of the toric codes when $N$ is more than two.} 
\par

Now we impose boundary condition on the lattice. We impose periodic boundary condition in both of $x$- and $y$-direction, assuming the length of the lattice in the $x$-direction is even. In this case, the Hamiltonian~\eqref{hamiltonian} consists of two decoupled $\mathbb{Z}_2$ toric codes on the torus. 
Such decomposition is portrayed in Fig.~\ref{sfb}.
In each $\mathbb{Z}_2$ toric code, there are two types of $\mathbb{Z}_2$ excitations, electric and magnetic charge. The non-local string of these charges yields logical operators, giving rise to fourfold GSD on the torus~\cite{KITAEV20032,surfacecode2012}. Therefore, GSD of our model is found to be $4^2=16$. On the contrary, the situation is drastically different when the length of the lattice in the $x$-direction is odd, instead of even. In this case, one cannot separate the terms~\eqref{ppa} into two groups, thus the decomposition~\eqref{group} is no longer true. Indeed, the terms belonging to (i) in~\eqref{group} are ``connected" with the ones belonging to (ii). 
For instance, the terms $V_{(n_x-1,y)}$ which belong to (i) in~\eqref{group} and $V_{n_x+1,y}=V_{1,y}$ belonging to (ii) (equality follows from the periodic boundary condition) are located to adjacent each other.
Therefore, the model is equivalent to one $\mathbb{Z}_2$ toric code on torus, giving $GSD=4$. 
\par
To summarize the argument presented in this subsection, in the simplest case by setting $N=2$, we learn that our model is identified as the $\mathbb{Z}_2$ toric code where each plaquette term is in a rhombus shape and that its GSD on torus, drastically changes depending on the length of the lattice in the $x$-direction, i.e.,
\begin{equation}
  \text{GSD}= \begin{cases}
16\;(n_x\;\text{even})\\
4\;(n_x\; \text{odd}).
     \end{cases}
\end{equation}
Such behavior can be understood by the fractional excitations with mobility constraint, analogously to the fracton topological phases. (See~\cite{ebisu2209anisotropic} for more details.) In the next subsection, we turn to the generic cases of $N$ and discuss its topological properties. \par

\subsection{Logical operators and Ground state degeneracy}\label{gsd}
In this subsection, we discuss our model in the generic cases of $N$. As opposed to the case with $N=2$, identifying the GSD and the logical operators is not so immediate in the generic case of~$N$. Thus, we employ an alternative approach to accomplish this task. We adopt an algebraic tool of the graph theory, which will play a pivotal role in the subsequent discussion on the entanglement entropy. 
As we mentioned below~\eqref{hamiltonian}, each term of the Hamiltonian involves the clock states in the next nearest neighboring as well as the nearest neighbor ones. Such feature can be succinctly described by the 
%
\textit{Laplacian matrix}(\textit{Laplacian}, in short), the graph theoretical analogue of the second order derivatives~\cite{chung1997spectral}.

\par
To this end, 
at each $y$ on the lattice, we define
the Laplacian $L$ as the $n_x\times n_x$ matrix indexed by vertices $(x,y)\;(1\leq x\leq n_x)\;\exists y$ running in the $x$-direction, which has the following form:
\begin{equation}
        L=\begin{pmatrix}
2 & -1&&& -1\\
-1 & 2&-1 &&\\
&-1&2&\ddots&\\
&&\ddots&\ddots&-1\\
-1&&&-1&2\label{lp}
\end{pmatrix}
\end{equation}
Equivalently, 
\begin{equation}
    (L)_{x,x^\prime}=\begin{cases}
        2\;(x=x^\prime)\\
        -1\;(x=x^\prime\pm1)\\
        0\;(\text{else}),
    \end{cases}\label{lap}
\end{equation}
where we have conventionally defined $x=n_x+1=1$ due to the periodic boundary condition.
The term $V_{(x,y)}$ at given $y$, constituting the Hamiltonian~\eqref{pp} now can be rewritten in terms of the matrix element of the Laplacian~\eqref{lap} via
\begin{equation}
V_{(x,y)}=X_{2,(x,y+1/2)}X_{2,(x,y-1/2)}^{\dagger}X_{1,(x,y)}^{-(L)_{x,x}}X_{1,(x+1,y)}^{-(L)_{x+1,x}}X_{1,(x-1,y)}^{-(L)_{x-1,x}},\label{ppv}
\end{equation}
where we conventionally set $X_{1,(x,y)}^{-1}=X_{1,(x,y)}^{\dagger}$.
Likewise, each term of $P_{(x,y+1/2)}$ can be rewritten in the similar manner as
\begin{equation}
P_{(x,y)}=Z_{1,(x,y+1)}Z_{1,(x,y)}^{\dagger}Z_{2,(x,y+1/2)}^{-(L)_{x,x}}Z_{2,(x-1,y+1/2)}^{-(L)_{x-1,x}}Z_{2,(x+1,y+1/2)}^{-(L)_{x+1,x}}.\label{ppv2}
\end{equation}

\begin{figure}
    \begin{center}
      \begin{subfigure}{0.84\textwidth}
  \includegraphics[width=\textwidth]{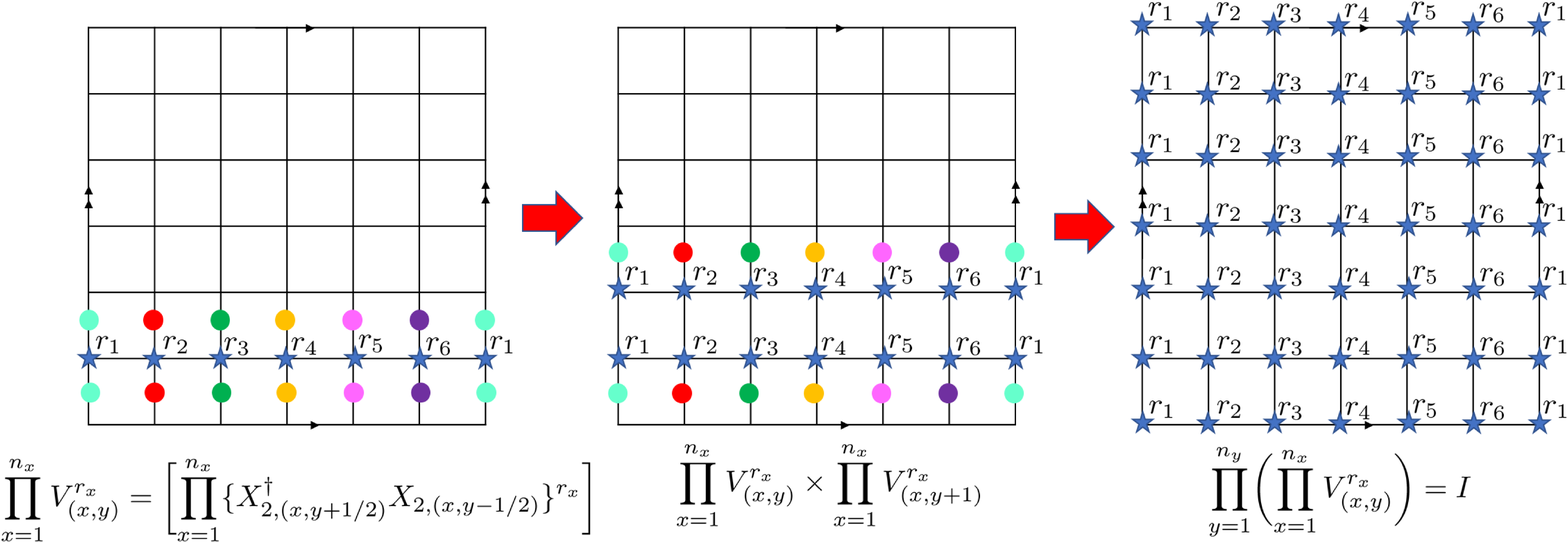}
         \caption{}\label{null}
             \end{subfigure}
        \begin{subfigure}{0.66\textwidth}
    \includegraphics[width=\textwidth]{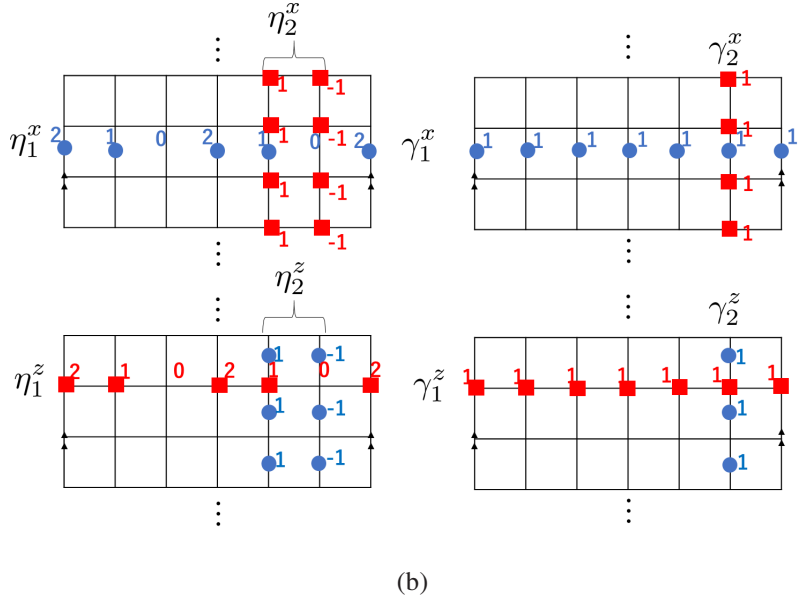}
         \caption{}\label{logical}
             \end{subfigure} 
\end{center}
\caption{(a) [Left] Configuration of the product of the operators $V_{(x,y)}$ (blue stars) given in~\eqref{cons},~$\prod_{x=1}^{n_x}V_{(x,y)}^{r_x}$ with $\mathbf{s}=\mathbf{0}$. Since $\mathbf{s}=\mathbf{0}$, it does not have $X_1$ operators in the horizontal direction, leaving $X_2$'s in the vertical links (dots). [Middle] If we multiply the product~\eqref{cons} with the adjacent one in the $y$-direction, some of $X_2$'s are cancelled out. [Right] Iterating the similar procedure, one obtains the product of~\eqref{cons} all along in the $y$-direction, $\prod_{y=1}^{n_y}\biggl(\prod_{x=1}^{n_x}V_{(x,y)}^{r_x}\biggr)$, which is identity due to the periodic boundary condition.
(b)[Top]~Configurations of the logical operators constructed by non-contractible loops of $X_1$ or $X_2$ operators defined in~\eqref{logi} in the case of $N=3$ and $n_x=6$. [Bottom]~Logical operators defined in~\eqref{logiz} with $N=3$ and $n_x=6$. The periodic boundary condition is imposed so that the left and right edge are identified. }
\label{fig:logical}
   \end{figure}
With this preparation, we now turn to counting the GSD on the torus geometry. The GSD is obtained by the number of total clock states divided by the number of independent stabilizers. 
The number of the clock states is found to be $N^{2n_xn_y}$. Also, the total number of the stabilizers~\eqref{pp} is given by $N^{2n_xn_y}$. However, we have over-counted the number of stabilizers; we
need to take into account the constraint on the stabilizers as multiplication of some of stabilizers becomes identity. To find such constraint, we resort to the formalism of the Laplacian~\eqref{lp}. At given $y$, introducing~$n_x$ dimensional vector $\mathbf{r}\vcentcolon=(r_1,\cdots,r_{n_x})^T\in\mathbb{Z}_N^{n_x}$, we  
consider following product of the vertex operators in the $x$-direction
\begin{equation*}
    \prod_{x=1}^{n_x}V_{\cc{x}{y}}^{r_x}.
\end{equation*}
Referring to~\eqref{ppv}, we transform it by using the Laplacian~$L$ as 
\begin{equation}
    \prod_{x=1}^{n_x}V_{\cc{x}{y}}^{r_x}=\biggl[\prod_{x=1}^{n_x}\{X^{\dagger}_{2,\cc{x}{y+1/2}}X_{2,\cc{x}{y-1/2}}\}^{r_x} \biggr]\times \prod_{x=1}^{n_x} X_{1,\cc{x}{y}}^{s_x}\;(s_x\in\mathbb{Z}_N)\label{cons}
\end{equation}
with 
\begin{equation}
    \mathbf{s}\vcentcolon=(s_1,\cdots,s_{n_x})^T=-L\mathbf{r}.
\end{equation}
Suppose $\mathbf{s}=\mathbf{0}\;( \text{mod} N)$. Then, multiplying the operators~\eqref{cons} along the $y$-direction gives (see also Fig.~\ref{null})
\begin{equation}
    \prod_{y=1}^{n_y}\biggl(\prod_{x=1}^{n_x}V_{\cc{x}{y}}^{r_x}\biggr)=\prod_{y=1}^{n_y}\biggl[\prod_{x=1}^{n_x}\{X^{\dagger}_{2,\cc{x}{y+1/2}}X_{2,\cc{x}{y-1/2}}\}^{r_x}\biggr]=I,
\end{equation}
where the last equation holds due to the periodic boundary condition.
Therefore, to find constraints of the stabilizers, we need to evaluate the solution of 
\begin{equation}
    L\mathbf{r}=\mathbf{0}\mod N, \label{kernel}
\end{equation}
i.e., the kernel of the Laplacian.\par
To proceed, we transform the Laplacian into the diagonal form, known as the \textit{Smith normal form}. Introducing invertible integer matrices $P$ and $Q$, one can transform the Laplacian~\eqref{lp} as
\begin{equation}
    PLQ=D,\;\;D=\text{diag}(\underbrace{1,\cdots,1}_{n_x-2},n_x,0),\label{snf}
\end{equation}
With this transformation, one can rewrite~\eqref{kernel} as
\begin{eqnarray}
\eqref{kernel}\Leftrightarrow P^{-1}DQ^{-1}\mathbf{r}=\mathbf{0}\mod N\nonumber\\\Leftrightarrow
   D \tilde{\mathbf{r}}=\mathbf{0}\mod N. 
   \label{con20}
   \end{eqnarray}
When moving from the second to the third equation, we have used the fact that $P$ is the integer matrix. Also, we have defined 
$\tilde{\mathbf{r}}\vcentcolon=Q^{-1}\mathbf{r}$. From~\eqref{snf}~\eqref{con20}, it follows that 
\begin{equation}
    \tilde{r}_i=0\;(1\leq i\leq n_x-2),\;\;n_x\tilde{r}_{n_x-1}=0\mod N.\label{con}
\end{equation}
The last entry of the vector $\tilde{r}$, $\tilde{r}_{n_x}$ takes $N$ distinct values since the last diagonal element of the matrix $D$ is zero~[\eqref{snf}]. Taking this fact as well as \eqref{con} into account, we have
\begin{equation}
    \mathbf{\tilde{r}}=(\underbrace{0,\cdots,0}_{n_x-2},N^\prime\alpha_1,\alpha_2)^T,\;N^\prime\vcentcolon=\frac{N}{\gcd(N,n_x)},\;(\alpha_1\in\mathbb{Z}_{\gcd(N,n_x)},\;\alpha_2\in\mathbb{Z}_N).
\end{equation}
Here, gcd stands for the greatest common divisor.
By evaluating the form of the matrix $Q$~\cite{ebisu2209anisotropic}, one finds
\begin{equation}
    \mathbf{r}=Q\tilde{\mathbf{r}}=
    N^\prime \alpha_1\begin{pmatrix}
    n_x-1\\n_x-2\\\vdots\\1\\0
    \end{pmatrix}
    +\alpha_2\begin{pmatrix}
    1\\1\\\vdots\\1\\1
    \end{pmatrix}\mod N. \label{kini}
\end{equation}
Therefore, there are $\gcd(N,n_x)\times N$ constraints on the stabilizers $V_{\cc{x}{y}}$. \par
One can analogously discuss the number of the constraints on the stabilizers $P_{\cc{x}{y+1/2}}$, yielding that there are $\gcd(N,n_x)\times N$ constraints. Overall, there are $[\gcd(N,n_x)\times N]^2$ constraints on the stabilizers, hence, the GSD is given by
\begin{equation}
    GSD=\frac{N^{2n_xn_y}}{N^{2n_xn_y}/[\gcd(N,n_x)\times N]^2}=[\gcd(N,n_x)\times N]^2.
\end{equation}
It is emphasized that our model exhibits unusual GSD dependence on the system size which was not seen in fracton topological phases which exhibit the sub-extensive GSD.\footnote{It is known that the Wen's $\mathbb{Z}_2$ plaquette model exhibits the similar gcd dependence of the GSD~\cite{wen2003}. However, our model discussed in the present paper is qualitatively different from the one in~\cite{wen2003} as our model admits the dipole of the fractional charges due to the multipole symmetries.}

One can make use of the formalism of the Laplacian to identify the form of the logical operators. By evaluating the kernel and cokernel of the Laplacian (see~\cite{ebisu2209anisotropic} for derivation), one can introduce logical operators consisting of non-contractible loop of $X_{1}$ or $X_2$ operators as
\begin{eqnarray}
    \eta^x_{1}=\prod_{x=1}^{n_x}X_{2,(x,y+1/2)}^{p_{x}},\; \gamma^x_{1}=\prod_{x=1}^{n_x}X_{2,(x,y+1/2)}^{q_{x}},\;\nonumber\\
    \eta^x_{2}=\bigl(\prod_{y=1}^{n_y}X_{1,(n_x-1,y+1/2)}\bigr)\times\bigl(\prod_{y=1}^{n_y}X^{\dagger}_{1,(n_x,y+1/2)}\bigr),\; \gamma^x_{2}=\prod_{y=1}^{n_y}X_{1,(n_x,y+1/2)}\label{logi}
    \end{eqnarray}
    with
    \begin{equation}
       \mathbf{p}=(p_1,p_2,\cdots,p_{n_x})^T= N^\prime \alpha_1\begin{pmatrix}
    n_x-1\\n_x-2\\\vdots\\1\\0
    \end{pmatrix},   \mathbf{q}=(q_1,q_2,\cdots,q_{n_x})^T=\alpha_2\begin{pmatrix}
    1\\1\\\vdots\\1\\1
        \end{pmatrix}\mod N\label{ppo}
    \end{equation}
    We portray them in Fig.~\ref{logical} in the case of $N=3$ and $n_x=6$.
    These operators are represented as $\eta^x_{m},\gamma_{m}^x\;(m=1,2)$, where two types of the loop are labeled by $\eta$ and $\gamma$, whereas the index $m$ denotes the direction of the loops ($x$ or~$y$). 
In addition to the logical operators, $\gamma^x_1$, $\gamma^x_2$ that one can find in the toric code, there are new forms of the logical operators in our model, $\eta^x_1$, $\eta^x_2$. In the horizontal direction, $\eta^x_1$ is formed by the inhomogeneous string of $X_2$'s whereas in the vertical direction, the model admits the ``dipole of the loops", i.e., a pair of the non-contractible loops of $X_1$'s in the vertical direction with opposite sign located adjacent to each other, giving $\eta^x_2$.  
The logical operators $\eta_{1}^x$ and $\gamma_{1}^x$ are topological in that they can be deformed upward or downward in the $y$-direction. 
    Also, logical operators 
    $\eta_{2}^x$ and $\gamma_{2}^x$ are deformable in the $x$-direction.\par
    
  Analogously to~\eqref{logi}, logical operators involving $Z_1$ or $Z_2$ operators are defined by (Fig.~\ref{logical})
    \begin{eqnarray}
 \eta^z_{1}=\prod_{x=1}^{n_x}Z_{1,(x,y)}^{p_{x}},\; \gamma^z_{1}=\prod_{x=1}^{n_x}Z_{1,(x,y)}^{q_{x}},\nonumber\\
     \eta^z_{2}=\bigl(\prod_{y=1}^{n_y}Z_{2,(n_x-1,y)}\bigr)\times\bigl(\prod_{y=1}^{n_y}Z^{\dagger}_{2,(n_x,y)}\bigr),\; \gamma^z_{2}=\prod_{y=1}^{n_y}Z_{2,(n_x,y)}, \label{logiz}
\end{eqnarray}
where the integers $p_x$ and $q_x$ are determined by~\eqref{ppo}.
It is straightforward to check that 
\begin{eqnarray}
    \eta^x_{1}\eta_{2}^z=\omega^{N^\prime}\eta_{2}^z \eta^x_{1},\;\gamma^x_{1}\gamma_{2}^z=\omega\gamma_{2}^z \gamma^x_{1},\;\eta^x_{2}\eta_{1}^z=\omega^{N^\prime}\eta_{1}^z \eta^x_{2},\;\gamma^x_{2}\gamma_{1}^z=\omega\gamma_{1}^z \gamma^x_{2}\nonumber\\
     (\eta^x_{1})^{\gcd(N,n_x)}=(\eta^x_{2})^{\gcd(N,n_x)}= (\eta^z_{1})^{\gcd(N,n_x)}=(\eta^z_{2})^{\gcd(N,n_x)}=I,\;\nonumber \\ (\gamma^x_{1})^{N}=(\gamma^x_{2})^{N}= (\gamma^z_{1})^{N}=(\gamma^z_{2})^{N}=I
    \label{cm}
\end{eqnarray}
with commutation relation between other combinations of the logical operators being trivial.
Logical operators with the commutation relation~\eqref{cm} yields the GSD $[\gcd(N,n_x)\times N]^2$. \par
Ground states with such GSD can be constructed by the logical operators. Labeling stabilizers involving $X_1$ or $X_2$ operators that 
are obtained by the product of vertex operators $V_{(x,y)}$
as 
\begin{equation*}
    G=\{g|g\in\prod_{x,y}V_{(x,y)}^{a_{x,y}},\;a_{x,y}\in\mathbb{Z}_N\},
\end{equation*}
we define the following ground state
\begin{equation}
   \ket{\psi}\vcentcolon= \frac{1}{\sqrt{|G|}}\sum_{g\in G}g\ket{0}
\end{equation}
where $\ket{0}$ is the trivial state in the diagonal basis of with the $Z_1$ and $Z_2$ operators, satisfying
$Z_{1,(x,y)}\ket{0}=Z_{2,(x,y+1/2)}\ket{0}=\ket{0}$ (the generalization of the ``spin-up state" to the clock state). Also, $|G|$ denotes the total number of the stabilizers $G$.
One can verify that this state is the stabilized state. The $[\gcd(N,n_x)\times N]^2$ ground states are labeled by 
\begin{equation}
\ket{\xi_{ab,cd}}\vcentcolon=(\eta^x_{1})^a(\gamma^x_{1})^b(\eta^x_{2})^c(\gamma^x_{2})^d\ket{\psi}\;\;(0\leq a,c\leq \gcd(N,n_x)-1, 0\leq b,d\leq N-1).\label{gs}
\end{equation}
It is known that the number of distinct fractional excitations in a topological phase on torus geometry is identical to GSD~\cite{Elitzur:1989nr}. Moreover, in our model, all of the fractional excitations are Abelian, implying that quantum dimension $d_a$ of any excitation is one. Taking these into account, one can evaluate the total quantum dimension as 
\begin{equation}
    \sqrt{\sum_ad^2_a}=N\times\gcd(N,n_x),\label{qd}
\end{equation}
where we have used the fact that there are $[N\times\gcd(N,n_x)]^2$ distinct fractional excitations, all of which carry quantum dimension one.

In the next section, we calculate the entanglement entropy of our model with respect to the ground states. 
In particular, we investigate whether the total quantum dimension~\eqref{qd} enters in the form of the entanglement entropy in the disk geometry.
\section{Entanglement entropy}\label{sec3}
Now we come to the main part of this paper. In this section, we study entanglement entropy of our model defined in the previous section. Our calculations rely on the formulation of the entanglement entropy in the stabilizer codes, invented in~\cite{Hamma2005}. 
\subsection{Review of the bipartite entanglement in lattice spin systems}
Before we systematically discuss the entanglement entropy of the higher rank topological phases, let's review the formalism of the bipartite entanglement in lattice spin systems proposed in~\cite{Hamma2005}. 
Readers familiar with this formalism may skip this subsection. 
\subsubsection{Reduced density matrix}
For a qubit system, the Hilbert space is $\m{H} = \otimes_j \m{H}_j$, where $\m{H}_j = \text{span}\{ |0\rangle_j, |1\rangle_j\, \dots, |N-1\rangle_j\}$ in the $\s^z_j$ basis, where $\s^z_j$ is generalized Pauli matrix in the $N$-clock model. We define a reference state $|0\rangle \vcentcolon= \otimes_j |0\rangle_j$, namely, all-spin-up state. We define $E \vcentcolon= \otimes_j \{1,\s^x_j, \dots, (\s^x_j)^{N-1}\}$ to be the Abelian group that rotates spins in an onsite manner. Any state in the Hilbert space can be written as $g|0\rangle$ for some $g \in E$ with $g^N = 1$. A generic state can be written as 
\begin{align}
|\psi\rangle &= \sum_{g \in G \subset E} a_g g|0\rangle, \ \sum_g |a_g|^2 = 1. \label{eqn:generic}
\end{align}
Here $G$ is a subgroup of $E$ and $a_g$ is the wavefunction of $|\psi\rangle$ in the computational basis, namely, $a_g = \langle 0 | g |\psi\rangle$. Given the state $|\psi\rangle$, we can define the corresponding density matrix as
\begin{align}
\rho_\psi &\vcentcolon= |\psi\rangle \langle \psi| = \sum_{g,g'} a_g \bar{a}_{g'} g|0\rangle \langle 0|g'^\dagger = \sum_{g,g'} a_g \bar{a}_{gg'} g|0\rangle \langle 0|(gg')^\dagger,
\end{align}
where in the last equality we redefined $g' \to gg'$. Once a bipartition of the lattice into subsystem A and its complement B, any element $g$ can be unambiguously decomposed into the product form $g = g_A \otimes g_B$, where $g_{A/B}$ only acts nontrivially on subsystems A/B, respectively.  With the decomposition of the reference state $|0\rangle = |0\rangle_A \otimes |0\rangle_B$, we obtain the reduced density matrix of subsystem A by tracing over subsystem B as
\begin{align}
\rho_A &= \sum_n {}_B\langle n | \rho |n \rangle_B = \sum_n \sum_{g,g'} a_g \bar{a}_{g g'} g_A |0\rangle_A {}_A \langle 0| (g_A g'_A)^\dagger \cdot {}_B \langle n| g_B |0\rangle_B {}_B\langle 0|(g_B g'_B)^\dagger |n \rangle_B \nonumber \\
&= \sum_{g,g'} a_g \bar{a}_{gg'} g_A |0\rangle_A {}_A \langle 0| (g_A g'_A)^\dagger \cdot {}_B \langle 0| g'^\dagger_B |0\rangle_B, 
\end{align}
where we used the completeness relation $\sum_n |0\rangle_B {}_B \langle 0| = I$. 
It is easy to see that nonzero contributions only come from $g'_B = I_B$. Define 
\begin{align}
G_A &\vcentcolon= \{ g \in G | g = g_A \otimes I_B \}, \ G_B \vcentcolon=\{ g \in G|g = I_A \otimes g_B\}.
\end{align}
Then $g'_B = I_B$ implies $g' \in G_A$. The reduced density matrix is rewritten as
\begin{align}
\rho_A &= \sum_{g \in G, g' \in G_A} a_g \bar{a}_{gg'} g_A |0\rangle_A {}_A \langle 0| (g_A g'_A)^\dagger. \label{eqn:reducedrho}
\end{align}
The equivalences between the following statements are proven~\cite{Hamma2005} : 1).~$\rho_A$ is diagonal, 2).~$G_A = \{ 1 \}$, 3).~$\nexists g = g_A \cdot g_B$ with both $g_A \in G_A, g_B \in G_B$ nontrivial for $g \in G$. 

\subsubsection{ Entanglement entropy of a subsystem}
We will study the entanglement entropy of subsystem A: 
\begin{align}
S_A &= - \text{tr~} \rho_A \text{log } \rho_A.
\end{align}
Without loss of generality, we can assume the total density matrix $\rho$ is a pure state. 
Let $S$ be the stabilizer group, which contains mutually commuting operators formed from $G$ that are called {\it stabilizers}. The codeword space is defined as $\m{L} = \{ |\psi\rangle \in \m{H} |~U |\psi\rangle = |\psi\rangle, \forall~U \in S\}$. Let us take $|\psi\rangle = \frac{1}{\sqrt{|G|}} \sum_{g \in G} g |0\rangle$ for the moment. We will study the generic ground states in the Appendix.~\ref{appendix:higherrankee}. Apparently, $|\psi\rangle \in \m{L}$ as $g |\psi\rangle = |\psi\rangle, \forall~ g \in G$. From~(\ref{eqn:reducedrho}), we have
\begin{align}
\rho_A &= \frac{1}{|G|} \sum_{g \in G, g' \in G_A} g_A |0\rangle_A {}_A \langle 0|(g_A g'_A)^\dagger. \label{eqn:reduced2}
\end{align}
Let $G/G_B$ and $G_{AB} \vcentcolon= G/(G_A \cdot G_B)$ be quotients that contain elements that act freely on $A$ and $A \cup B$, respectively. It is shown that for the ground state~(\ref{eqn:reduced2}), the entanglement entropy is described by~\cite{Hamma2005}
\begin{align}
S_A &= \log |G_{AB}|. \label{eqn:EE}
\end{align}
The details are given in Appendix~\ref{appendix:EEproof}.
The entanglement entropy of the subsystem $A$ is given by
\begin{equation}
    S_A=\log|G|-\log|G_A|-\log|G_B|, \label{eqn:EEequalsuperposition}
\end{equation}
where $|G|$ denotes the total number of independent stabilizers containing $X_1$ and $X_2$, whereas $|G_A|$ ($|G_B|$) represents the number of  stabilizers containing $X_1$ and $X_2$ that have support on $A (B)$.

In the following subsections, we study the entanglement entropy of various geometries based on the formula~\eqref{eqn:EEequalsuperposition}. 

\subsection{Single-row/column geometry}\label{singlerowco}

\begin{figure}[h]
    \begin{center}
      \begin{subfigure}[h]{0.20\textwidth}
  \includegraphics[width=\textwidth]{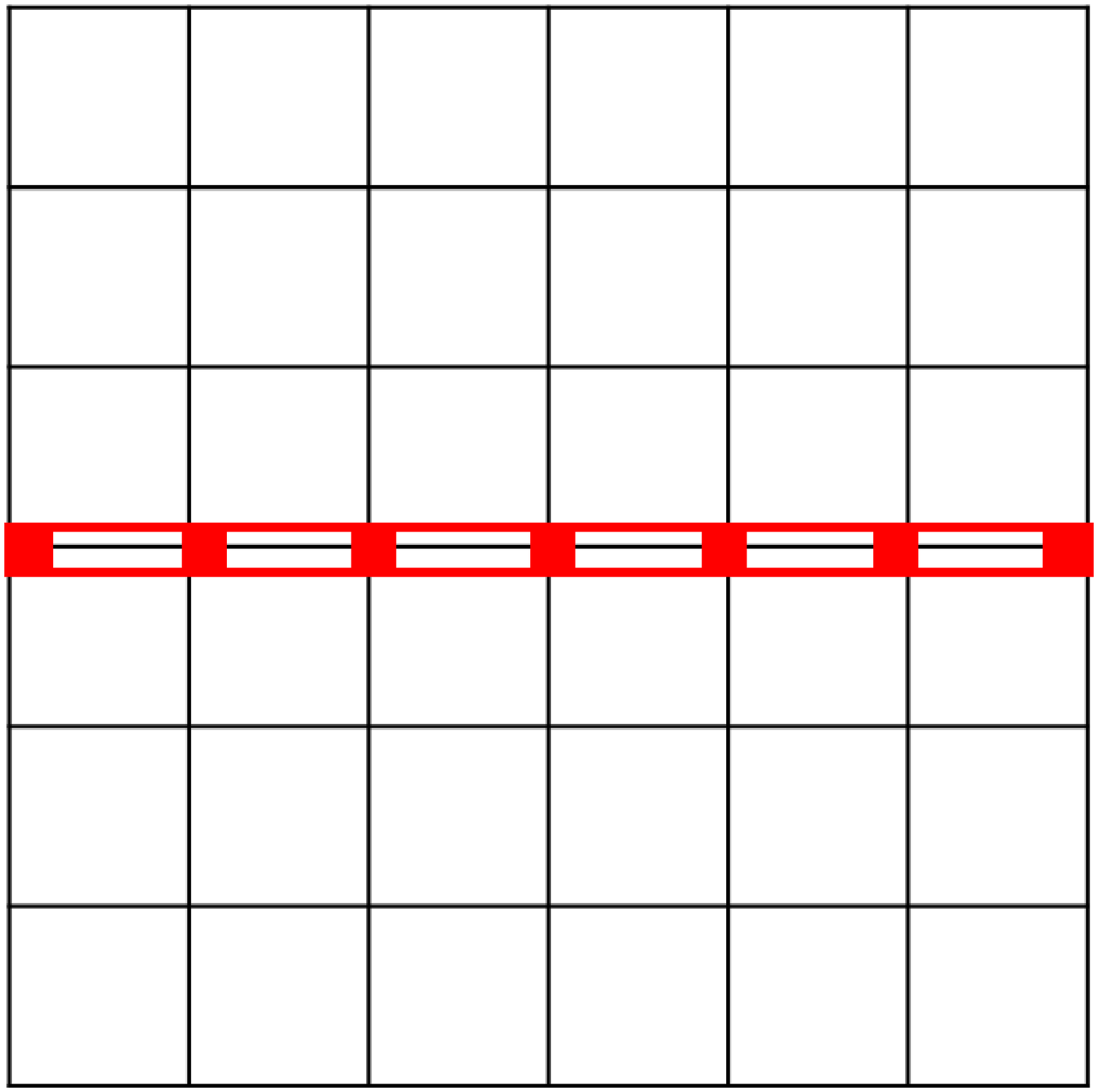}
         \caption{}\label{4a}
             \end{subfigure}
        \begin{subfigure}[h]{0.20\textwidth}
    \includegraphics[width=\textwidth]{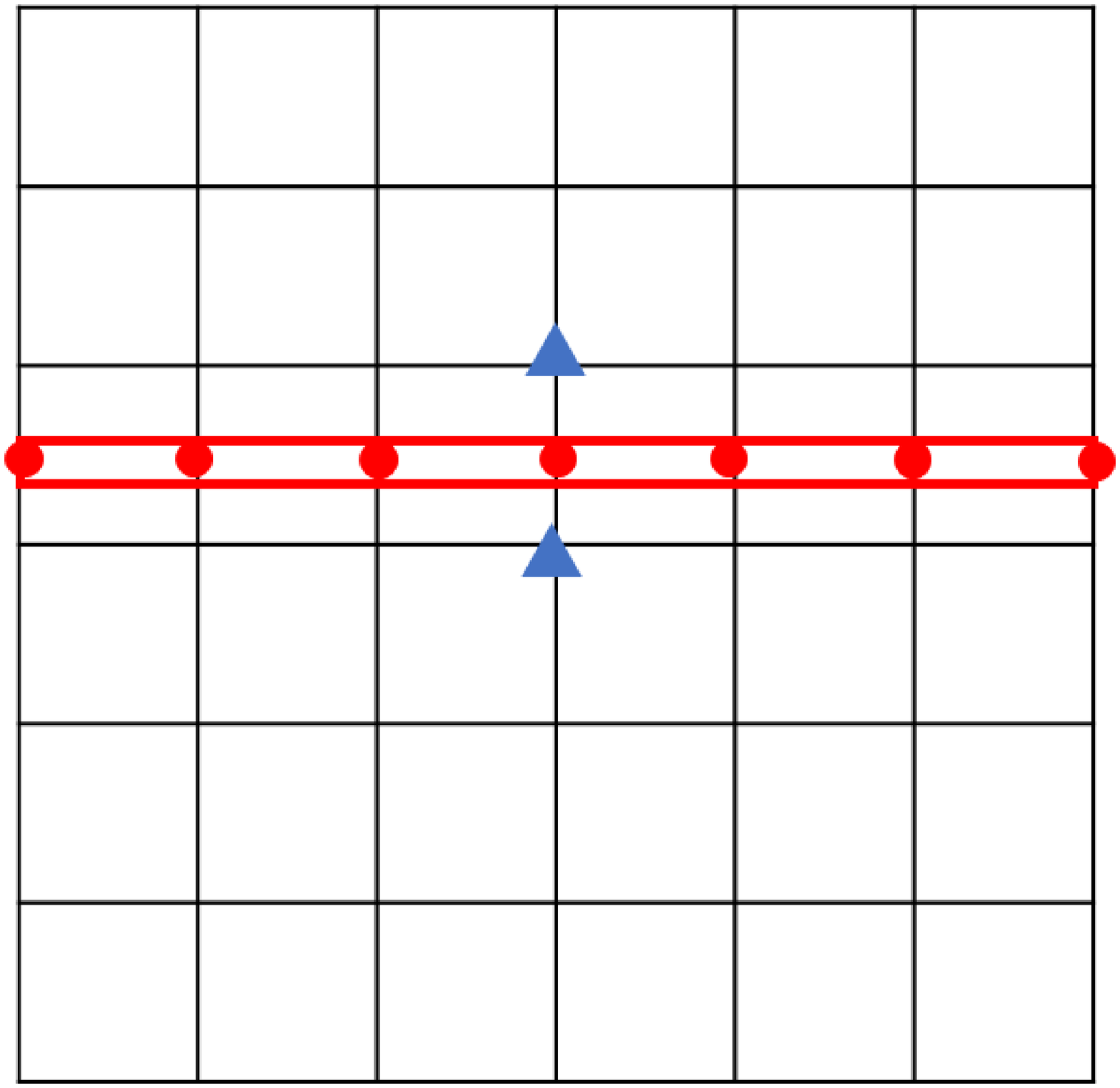}
         \caption{}\label{4b}
             \end{subfigure} 
                \begin{subfigure}[h]{0.20\textwidth}
    \includegraphics[width=\textwidth]{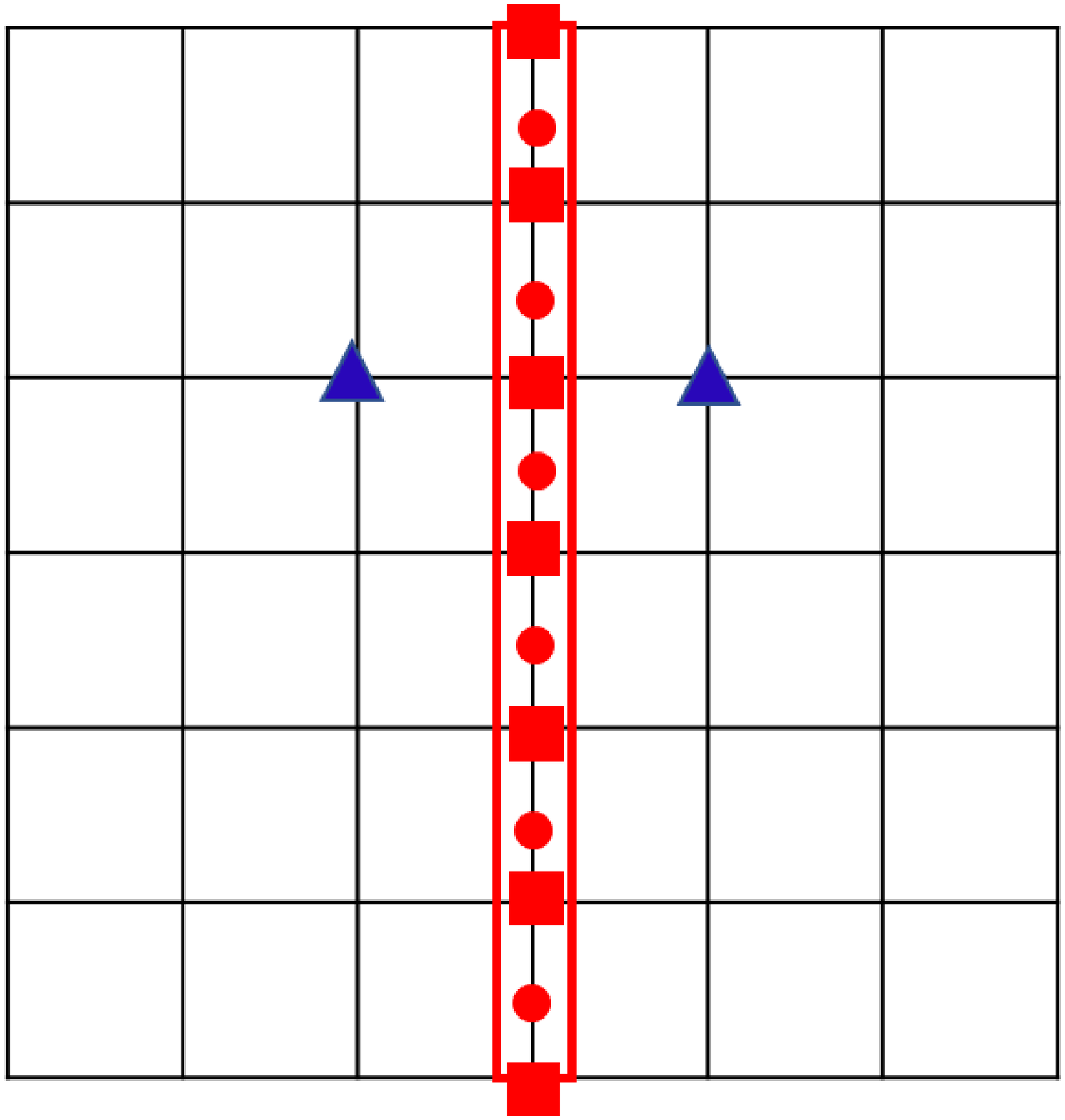}
         \caption{}\label{4c}
             \end{subfigure} 
              \begin{subfigure}[h]{0.20\textwidth}
    \includegraphics[width=\textwidth]{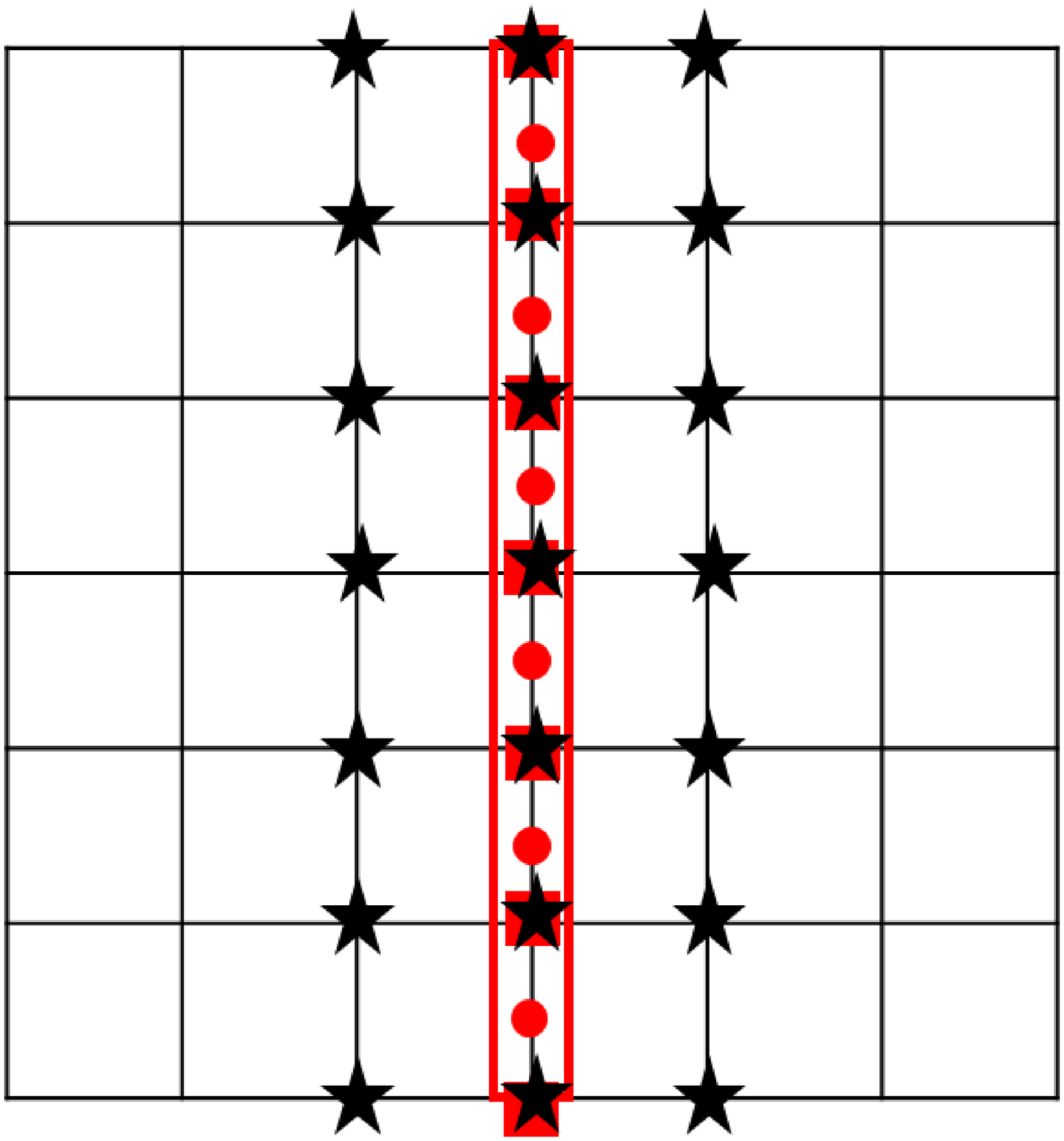}
         \caption{}\label{product}
             \end{subfigure} 
              \begin{subfigure}[h]{0.22\textwidth}
    \includegraphics[width=\textwidth]{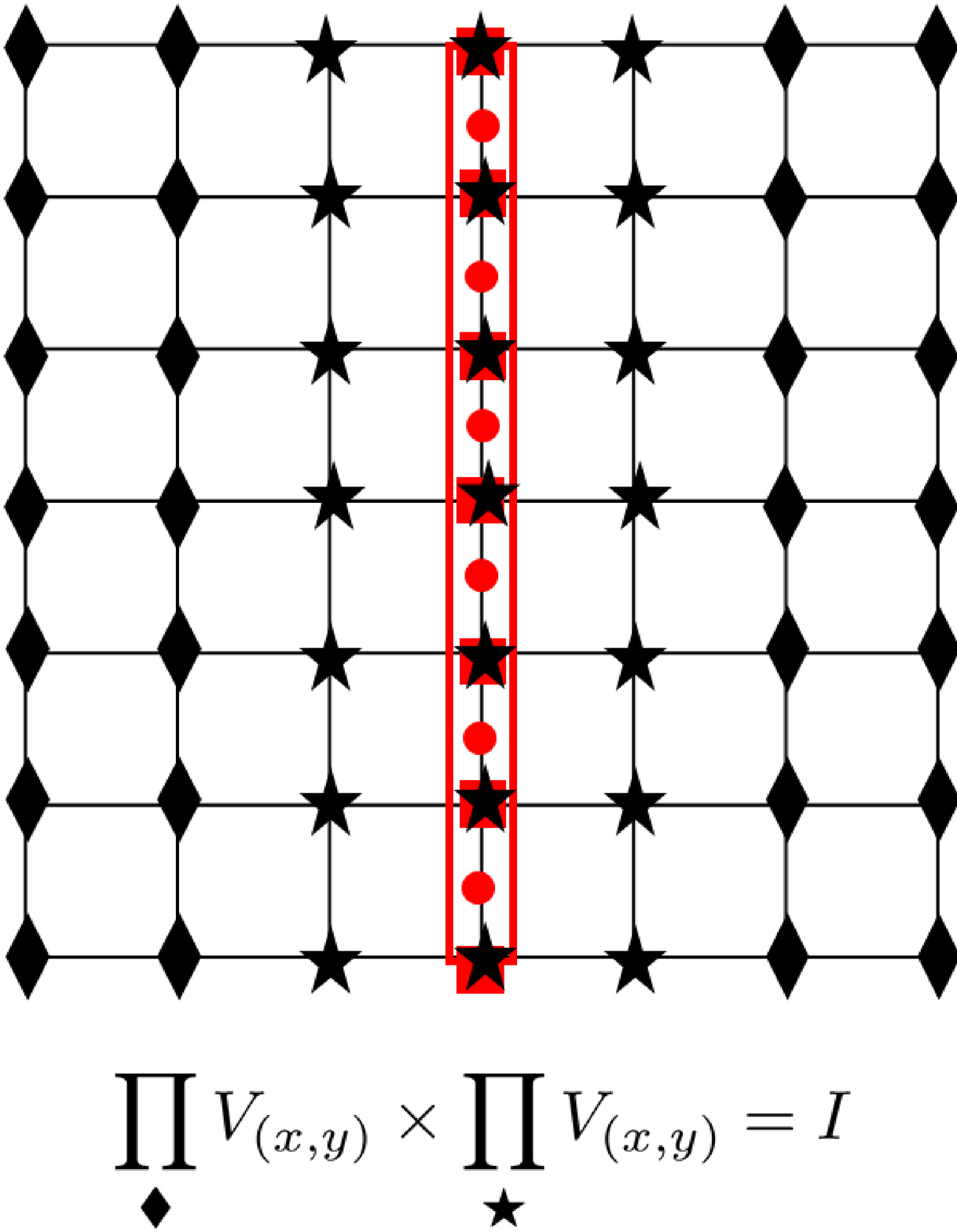}
         \caption{}\label{constraint}
             \end{subfigure} 
\end{center}
\caption{
(a)~The subsystem A is a single row on the horizontal sites (clock states inside the red frame). (b)~The subsystem A is a single row on the vertical links (red dots inside the red frame). The blue triangles represent one of the $n_x$ constraints along the horizontal direction. (c)~The subsystem A is a single column (red dots and squares inside the frame). The blue triangles represent one of the $n_y$ constraints along the vertical sites. 
(d)~Multiplication of the~vertex operators (represented by black stars) that act within $B$. (e)~Such product of the stabilizers is redundant due to the constraint that product of all of the~vertex operators becomes identity~\eqref{star}.
}
\label{fig:singlerow}
\end{figure}
After reviewing formulation of the entanglement entropy, we calculate it in various geometries of subsystems in our model. 
We evaluate the entanglement entropy with respect to the ground state $\ket{\xi_{00,00}}$ given in~\eqref{gs}. One could study the entanglement entropy for more generic ground state $\ket{\zeta}$ which is superposition of $\ket{\xi_{ab,cd}}$, defined by
\begin{equation}
\ket{\zeta}=\sum_{a,c=0}^{\gcd(N,n_x)-1}\sum_{b,d=0}^{N-1}\alpha_{ab,cd}\ket{\xi_{ab,cd}}\;\;\biggl(\sum_{abcd}|\alpha_{ab,cd}|^2=1,\;\alpha_{ab,cd}\in U(1)\biggr).\label{generic}
\end{equation}
In this case, 
depending on the geometry of the subsystem $A$, there is an additional contribution to~\eqref{eqn:EEequalsuperposition}.
While we focus on the entanglement entropy with respect to the ground state $\ket{\xi_{00,00}}$, we 
make a brief comment on the case of the generic ground state~\eqref{generic}, deferring the details to Appendix~\ref{appendix:higherrankee}.

In this subsection, we consider 
the single-row or-column subsystem geometries which go around the torus in the $x$- or $y$-direction, as shown in Fig.~\ref{fig:singlerow}. 
When calculating the entanglement entropy based on~\eqref{eqn:EEequalsuperposition}, there is a technical caveat: when evaluating $|G_A|$ and $|G_B|$, we need to take into account product of vertex operators which have support only on $A$ and $B$. 


\subsubsection{Single row I}
Let's first look at 
the subsystem A in Fig.~(\ref{fig:singlerow})(a), we have $|G|=N^{n_xn_y}/{\Gamma}$, 
$|G_A|=1$, $|G_B| = N^{n_xn_y-n_x}$, where the factor $\Gamma$ corresponds to the two independent constraints from~(\ref{kini}), $\Gamma\vcentcolon=N\times\gcd(N,n_x)$.
        Thus, (\ref{eqn:EEequalsuperposition}) gives us
\begin{align}
S_A &= n_x\log N-\log N-\log[\gcd(N,n_x)]=n_x\log N-\log\Gamma.\label{r1}
\end{align}
For the generic ground state (\ref{generic}), there is an additional constant $\mu_1$ in ~\eqref{r1}:
\begin{equation}
   S_A=n_x \log{N}-\log\Gamma+\mu_1,\;\; 
   \mu_1=-\sum_{c,d}\biggl[\biggl(\sum_{a,b}|\alpha_{ab,cd}|^2\biggl)\log\biggl(\sum_{a,b}|\alpha_{ab,cd}|^2\biggl) \biggr]\label{r1prime}
\end{equation}
Derivation of~\eqref{r1prime} is given in Appendix~\ref{appendix:higherrankee}. 

\subsubsection{Single row II}
For the subsystem A in Fig.~(\ref{fig:singlerow})(b), we have 
\begin{equation*}
    |G_A| = 1, |G_B| = (|G|/N^{2n_x}) \cdot N^{n_x} = |G|/N^{n_x},
\end{equation*}
When evaluating $|G_B|$, the $N^{n_x}$ in the first equality comes from the $n_x$ independent constraints, one of which is labeled by the blue triangles in Fig.~(\ref{fig:singlerow})(b). Following~(\ref{eqn:EEequalsuperposition}), we have 
\begin{align}
S_A &= n_x \log{N}.\label{r2}
\end{align}
In the case of the generic ground state~\eqref{generic}, we have an additional contribution to~\eqref{r2}:
\begin{equation}
S_A=n_x\log{N}+\mu_2,\;\;\mu_2=-\sum_{p=0}^{\gcd(N,n_x)-1}\sum_{q=0}^{N-1}\lambda_{p,q}\log\lambda_{p,q}\label{nu1}
\end{equation}
with ($\nu=e^{2\pi i/\gcd(N,n_x)}$, $\omega=e^{2\pi i/N}$)
\begin{equation}
    \lambda_{p,q}=\frac{1}{\Gamma}\sum_{k=0}^{\gcd(N,n_x)-1}\sum_{l=0}^{N-1}\nu^{pk}\omega^{ql}\sigma_{kl},\;\;\sigma_{kl}= \sum_{\substack{a,a^\prime,b,b^\prime,c,d \\ a-a^\prime=k\mod \gcd(N,n_x)\\b-b^\prime=l\mod N}}\alpha_{ab,cd}\bar{\alpha}_{a^\prime b^\prime,cd}.\label{nu2}
\end{equation}
The derivation of this constant is relegated to Appendix~\ref{appendix:higherrankee}. 

\subsubsection{Single column}
Let us turn to subsystem A in Fig.~(\ref{fig:singlerow})(c).
The total number of the independent vertex operators is given by 
 $|G|=N^{n_xn_y}/\Gamma$, 
where we take the constraint discussed in the previous section 
into consideration (Sec.~\ref{gsd}). Next we evaluate $|G_B|$. Naively, 
the number of stabilizers that act within $B$ is given by
\begin{equation}
N^{n_xn_y-3n_y}\label{44}
\end{equation}
However, this counting is incorrect, as
one has to take into account the multiplication of the stabilizers that act trivially on $A$. There are two types of such product. 
The first type is the product of the vertex operators on the both sides of the vertical line. One of such configuration is indicated by the blue triangles in Fig.~(\ref{fig:singlerow})(c).
There are $N^{n_y}$ such products. Another type is multiplication of the vertex operators which are located inside $A$ and the ones surrounding $A$, as shown in black stars in Fig.~(\ref{fig:singlerow})(d).
Having identified these product of the stabilizers that contribute to~$|G_B|$, one has to check whether these products are ``redundant" to 
the constraint of the stabilizes that we have already considered. 
\par
To be more specific to what we have just said, we look at the second type of the product of the stabilizers that are depicted in Fig.~\ref{fig:singlerow}(d). We schematically write this product as $\prod_{\bigstar}V_{(x,y)}$ in accordance with Fig.~\ref{fig:singlerow}(d).
As we discussed in Sec.~\ref{gsd}, the multiplication of all of the vertex operators becomes identity, i.e.,
\begin{equation}
    \prod_{\blacklozenge}V_{(x,y)}\times \prod_{\bigstar}V_{(x,y)}=I, \label{star}
\end{equation}
where $ \prod_{\blacklozenge}V_{(x,y)}$ denotes the product of the vertex operators defined on the complement of $\bigstar$ [see Fig.~\ref{fig:singlerow}(e)].
Hence, the product in question, $\prod_{\bigstar}V_{(x,y)}$
is redundant: it can be generated by the vertex operators belonging to $B$. 
Likewise, regarding the first type of the product, one of which 
is depicted in Fig.~(\ref{fig:singlerow})(c), combination of some of them is redundant due to the constraint.\par

Overall, the multiplication of the vertex operators that act on $B$ is found to be 
\begin{equation}
    \frac{N^{n_y}\times N}{[\gcd(N,n_x)\times N]}. 
\end{equation}
Therefore, $|G_B|$ is obtained by multiplying this number with~\eqref{44}, that is,
\begin{equation}
    |G_B|=N^{n_xn_y-3n_y}\times \frac{N^{n_y}\times N}{[\gcd(N,n_x)\times N]}.
\end{equation}
There is no stabilizer that act within $A$, giving $|G_A|=1$. Referring to~\eqref{eqn:EEequalsuperposition}, the entanglement entropy is given by
\begin{equation}
    S_A=2n_y\log N-\log N.
\end{equation}
In the case of the generic ground state~\eqref{generic}, 
\begin{equation}
    S_A=2n_y\log N-\log N+\mu_3,
   \;\; \mu_3=-\sum_{b=0}^{N-1}\sum_{q=0}^{N-1}\lambda_{q}^{(b)}\log\lambda_{q}^{(b)},\label{nu3}
\end{equation}
where
\begin{equation}
    \lambda_{q}^{(b)}=\frac{1}{N}\sum_{k=0}^{N-1}\omega^{kq}\sigma^{(b)}_k,\;\;\sigma^{(b)}_k=\sum_{\substack{a,c,d,d^\prime\\ d-d^\prime=k\mod N}}\alpha_{ab,cd}\bar{\alpha}_{ab,cd^\prime}.\label{nu4}
\end{equation}
The derivation is given in Appendix~\ref{appendix:higherrankee}.

\subsection{Disk geometry}\label{diskc}
\begin{figure}
    \begin{center}
       \begin{subfigure}[h]{0.34\textwidth}
  \includegraphics[width=\textwidth]{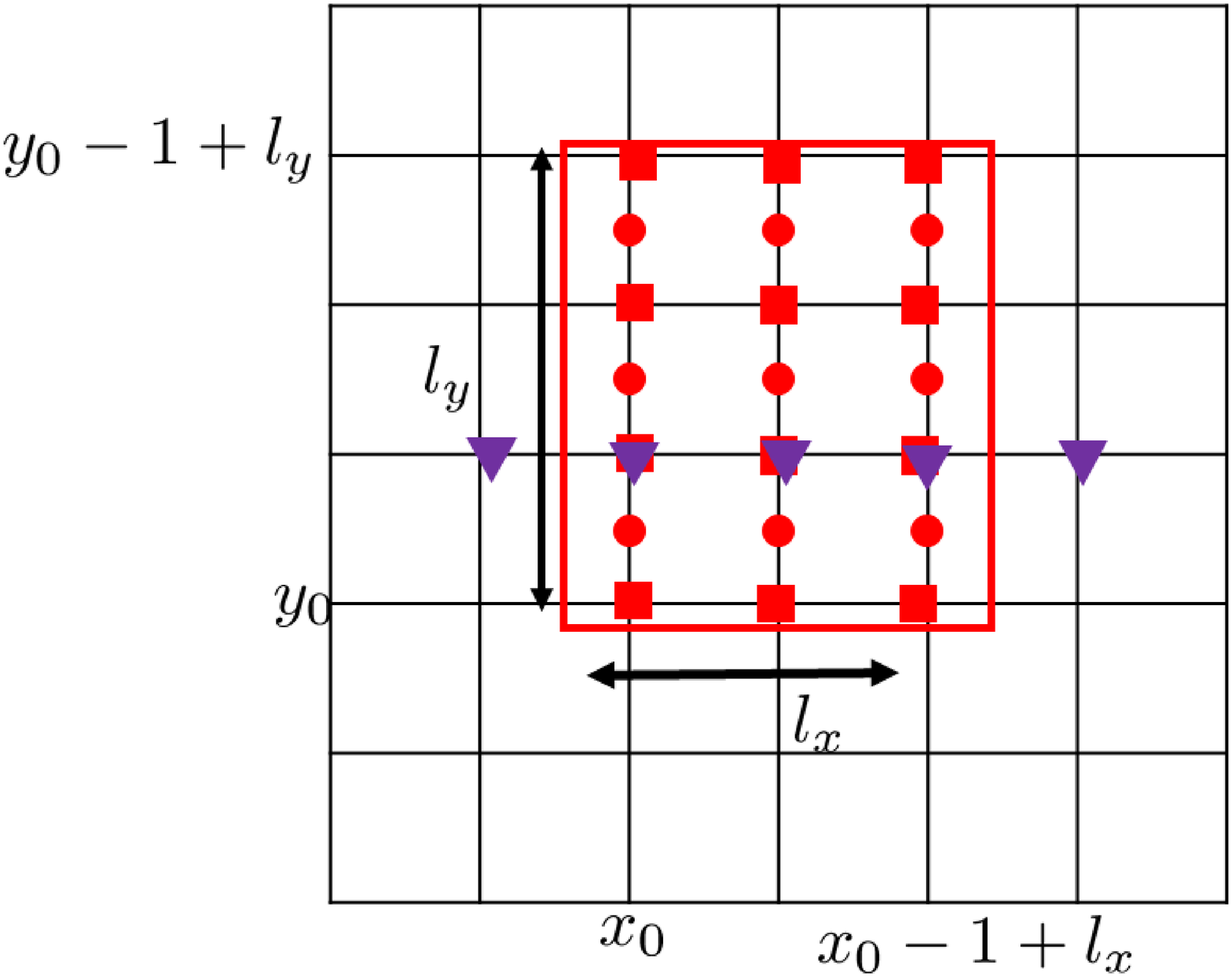}
         \caption{}\label{disk}
             \end{subfigure}
             \hspace{5mm}
               \begin{subfigure}[h]{0.59\textwidth}
    \includegraphics[width=\textwidth]{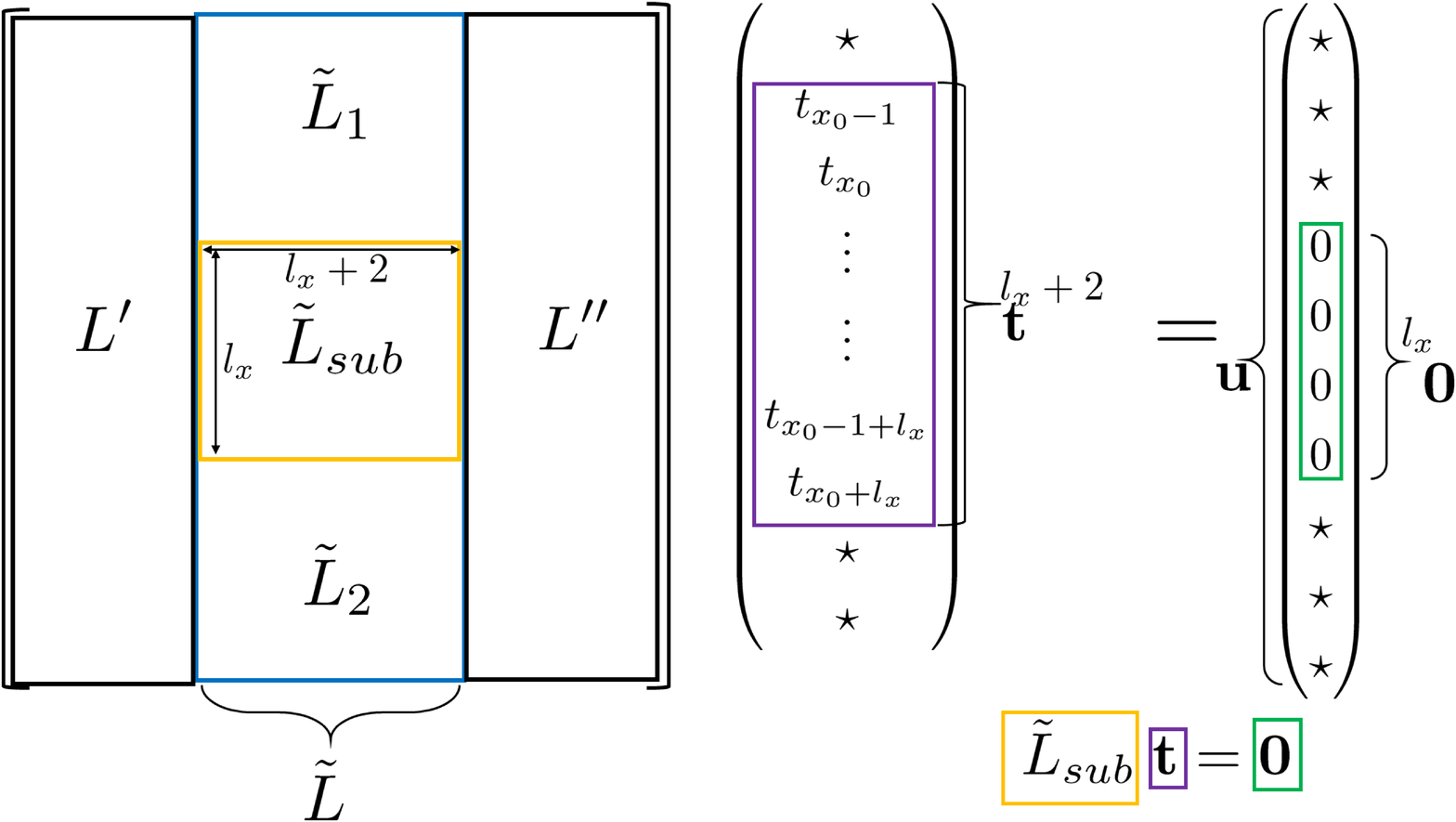}
         \caption{}\label{pd4}
             \end{subfigure} 
                  \begin{subfigure}[h]{0.39\textwidth}
    \includegraphics[width=\textwidth]{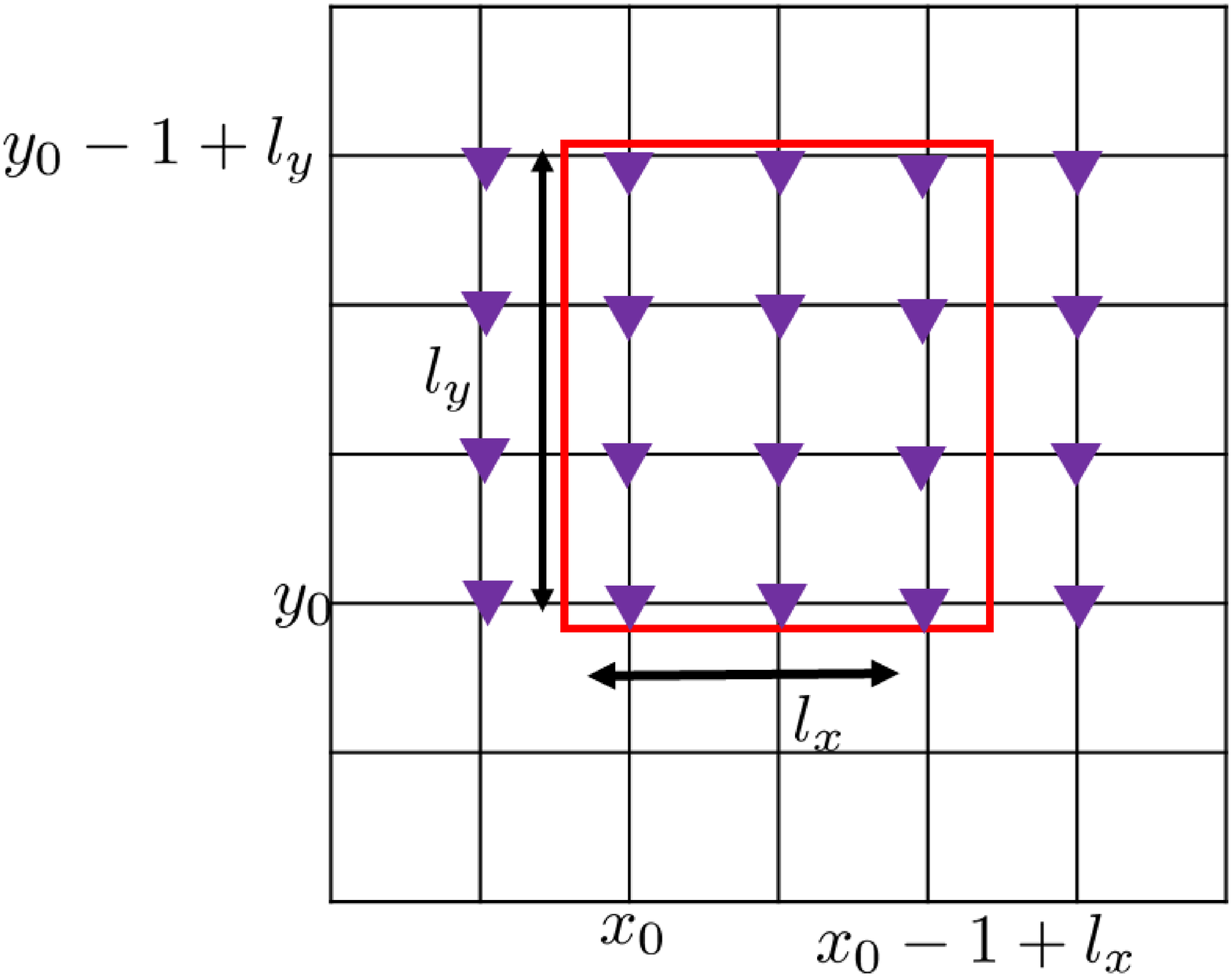}
         \caption{}\label{ote}
             \end{subfigure} 
             \hspace{5mm}
                    \begin{subfigure}[h]{0.29\textwidth}
    \includegraphics[width=\textwidth]{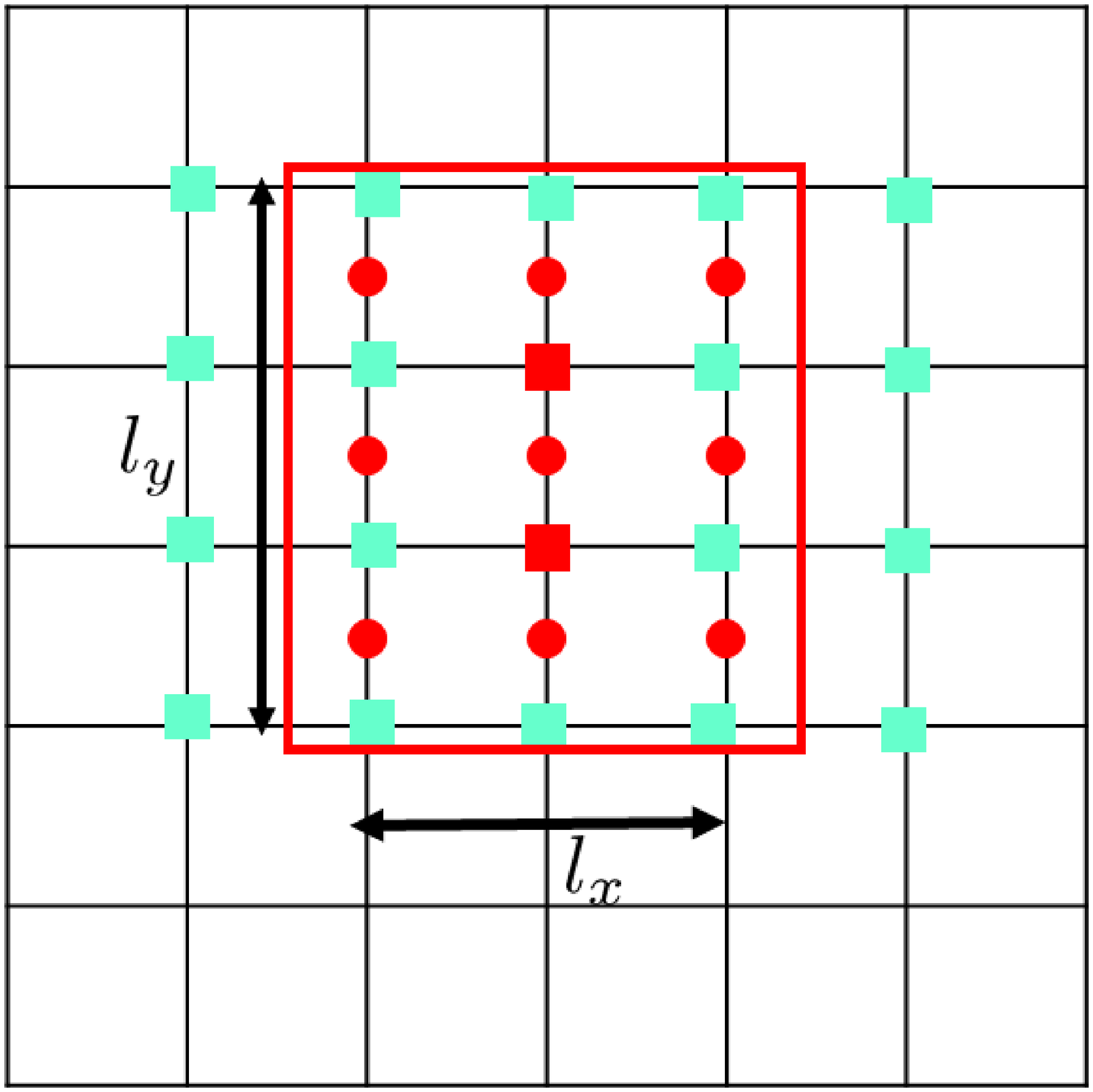}
         \caption{}\label{area}
             \end{subfigure} 
 \end{center}
 \caption{(a)~Disk geometry. The clock states inside the disk are marked by red squares and dots. Purple triangles represent the product given in~\eqref{shine}. (b) Decomposition of the Laplacian into sub-matrices, corresponding to~\eqref{lp222}~\eqref{lp2223}. We need to find multiplication of the vertex operators corresponding to the vector $\mathbf{t}$ which does not have $X_1$ operators inside the disk, yielding the condition~\eqref{ti}. (c) Configuration of the product given in~\eqref{hoo} (purple triangles). With the condition~\eqref{usol}, it exclusively acts on $B$. (d) The number of the clock states marked by light-blue squares gives rise to the area-law term, $(Area)$, given in~\eqref{main}, which is $(Area)=4l_y+2(l_x-2)$ in the present case. }
 \end{figure}

Let us calculate entanglement of the contractible disk geometry, as portrayed in Fig.~\ref{disk}. 
We assume that width of the disk is more than one.
We set the width and height of the disk to be $l_x (\geq 2)$ and $l_y$, respectively. Accordingly, the coordinate of $\mathbb{Z}_N$ clock states with type 1 (i.e., clock states located at vertices) inside 
the disk is denoted as $(x,y)\;(x_0\leq x\leq x_0+l_x-1, y_0\leq y\leq y_{0}+l_y-1)$ and the ones with type 2 (i.e, clock states at vertical links) inside the disk as $(x,y+1/2)\;(x_0\leq x\leq l_x-1+x_0, y_0\leq y\leq y_0+l_y-2)$. See also Fig.~\ref{disk}.

In this setting, we have $|G|=N^{n_xn_y}/\Gamma$, $|G_A|=N^{(l_x-2)(l_y-2)}$. As for $|G_B|$, the number of individual stabilizers that have support only on $B$ is
\begin{equation}
    N^{n_xn_y}/N^{l_y(l_x+2)}\label{oote}
\end{equation}
In addition to this number, we need to take the product of the vertex operators that act within $B$ into consideration. To find such product, it is useful to resort to the formalism of the Laplacian that we have discussed in Sec.~\ref{gsd}.
Consider the following product of $V_{(x,y)}$'s in the horizontal direction at given $y\;(y_0\leq y\leq y_{0}+l_y-1)$ (purple triangles in Fig.~\ref{disk})
\begin{equation}
   \prod_{x=x_0}^{x_0+l_x-1}V_{(x,y)}^{t_x}\;\;(t_x\in\mathbb{Z}_N),\label{shine}
\end{equation}
which can be rewritten as 
\begin{equation}
   \eqref{shine}=\biggl[\prod_{x=x_0}^{x_0+l_x-1}\{X^{\dagger}_{2,\cc{x}{y+1/2}}X_{2,\cc{x}{y-1/2}}\}^{t_x} \biggr]\times \prod_{x=1}^{n_x} X_{1,\cc{x}{y}}^{u_x}\;(u_x\in\mathbb{Z}_N).\label{ss}
\end{equation}
Analogously to the discussion in the previous section, Sec.~\ref{gsd}, $\mathbb{Z}_N$ numbers $t_x$ and $u_x$ that appear in~\eqref{shine}~\eqref{ss}
are related via Laplacian. 
Defining $l_x+2$- and $n_x$-dimensional vector by $\mathbf{t}\vcentcolon=(\underbrace{t_{x_0-1},\cdots,t_{x_0+l_x}}_{l_x+2})^T$, and $\mathbf{u}\vcentcolon=(\underbrace{u_{1},\cdots,u_{n_x}}_{n_x})^T$, respectively, 
and
referring to~\eqref{lp}~\eqref{lap}, 
we have
\begin{equation}
    \mathbf{u}=-\tilde{L}\mathbf{t}.\label{hi}
\end{equation}
    Here, $\tilde{L}$ represents a sub-matrix of the Laplacian $L$~\eqref{lap}. It is obtained by decomposing the Laplacian into three matrices as
\begin{equation}
    L=\begin{pmatrix}
 & & \\
 L^\prime&\tilde{L}&L^{\prime\prime}\\
&&\label{lp222}
\end{pmatrix},
\end{equation}
where $L^\prime$, $\tilde{L}$, $L^{\prime\prime}$ is $n_x\times (x_0-2)$, $n_x\times (l_x+2)$, $n_x\times (n_x-x_0-l_x)$ sub-matrix, respectively, and concentrating on the middle matrix, $\tilde{L}$ (see also Fig.~\ref{pd4}).

Suppose the product we consider~\eqref{ss} with~\eqref{hi} has no $X_1$ operators within the disk $A$, that is, 
\begin{equation}
    \mathbf{u}=(u_1,\cdots,u_{x_0-1},\underbrace{0,\cdots,0}_{l_x},u_{x_0+l_x},\cdots,u_{n_x})^T.\label{usol}
\end{equation}
Then, multiplying the product~\eqref{ss} along the $y$-direction gives
\begin{equation}
    \prod_{y=y_0}^{y_0+l_y-1}\biggl[\prod_{x=x_0}^{x_0+l_x-1}V_{(x,y)}^{t_x}\biggr]\label{hoo}
\end{equation}
which yields the stabilizers that have support only on $B$ (see Fig.~\ref{ote}). Therefore, to find the product of stabilizers that acts within $B$, we need to evaluate~\eqref{hi} with~\eqref{usol}. Such condition can be rewritten as 
\begin{equation}
    \tilde{L}_{sub}\mathbf{t}=\mathbf{0}, \label{ti}
\end{equation}
where $\tilde{L}_{sub}$ denotes the sub-matrix of $\tilde{L}$, which is obtained by decomposing the matrix $\tilde{L}$ into three via
\begin{equation}
    \tilde{L}=\begin{pmatrix}
 &\tilde{L}_1& \\
&\tilde{L}_{sub}&\\
&\tilde{L}_2&\label{lp2223}
\end{pmatrix},
\end{equation}
where $\tilde{L}_1$, $\tilde{L}_{sub}$, and $\tilde{L}_{2}$ is the $(x_0-1)\times(l_x+2)$, $l_x\times(l_x+2)$, and $(n_x-l_x-x_0+1)\times(l_x+2)$ matrix, respectively, and focusing on the middle matrix $\tilde{L}_{sub}$ (Fig.~\ref{pd4}).
From~\eqref{lp}~\eqref{lp222}~\eqref{lp2223}, the explicit form of~$\tilde{L}_{sub}$ is given by
\begin{equation}
    \tilde{L}_{sub}=\begin{pmatrix}
-1&2 & -1&&& &\\
&-1 & 2&-1 &&&\\
&&-1&2&\ddots&&\\
&&&\ddots&\ddots&-1&\\
&&&&-1&2&-1\label{lp3}
\end{pmatrix}.
\end{equation}
Introducing $l_x\times l_x$ and $(l_x+2)\times (l_x+2)$ invertible matrices, $\tilde{P}$ and $\tilde{Q}$, whose matrix elements are integers, 
one can transform the matrix~$\tilde{L}_{sub}$ into a diagonal form (a.k.a. the \textit{Smith normal form}) as $\tilde{P}\tilde{L}_{sub}\tilde{Q}=\tilde{D}$ with
\begin{equation}
    \tilde{D}=\begin{pmatrix}
1& & &&& &&&&\\
&1 & & &&&&&&\\
&&1&&&&
&&&\\
&&&\ddots&&&&&&\\
&&&&&&&&&\\
&&&&&&&1&0&0\label{snf22}
\end{pmatrix}.
\end{equation}
The $l_x\times(l_x+2)$~matrix $\tilde{D}$ contains diagonal entries $\text{diag}(\underbrace{1,\cdots,1}_{l_x})$ with other entries being zero.
With this form, it follows that 
\begin{equation}
    \tilde{L}_{sub}\mathbf{t}=\mathbf{0}\Leftrightarrow \tilde{D}\tilde{\mathbf{t}}=\mathbf{0}\mod N,\label{snf3}
\end{equation}
where $\mathbf{\tilde{t}}\vcentcolon=\tilde{Q}^{-1}\mathbf{t}$ and $\mathbf{0}$ represents $l_x$-dimensional zero vector. We rename the entries of the vector~$\mathbf{\tilde{t}}$ as~$\mathbf{\tilde{t}}=(\tilde{t}_1,\cdots,\tilde{t}_{l_x+2})^T$.
From the Smith normal form~\eqref{snf22} and~\eqref{snf3}, the first $l_x$  entries of the vector $\mathbf{\tilde{t}}$ are subject to
\begin{equation}
    \tilde{t}_i=0\;(1\leq i\leq l_x), 
\end{equation}
whereas there is no constraint on the last two entries $\tilde{t}_{l_x+1}$, $\tilde{t}_{l_x+2}$, giving
\begin{equation}
    \tilde{t}_{l_x+1}=\beta_1,\;  \tilde{t}_{l_x+2}=\beta_2\;(\beta_1,\beta_2\in\mathbb{Z}_N).
\end{equation}
Hence, there are $N^2$ solutions of~\eqref{ti}. \par
One has to check that such $N^2$ product of vertex operators which have support only on $B$ is redundant to the constraint of the stabilizers, analogously to the discussion presented around~\eqref{star}. 
Thus, the number of product of the stabilizers that act within $B$ is given by
$ \frac{N^2}{\Gamma}$.

Overall, we obtain $|G_B|$ by multiplying this number with~\eqref{oote}:
\begin{equation}
    |G_B|=\frac{N^{n_xn_y}}{N^{l_x(l_y+2)}}\times\frac{N^2}{\Gamma}.
\end{equation}
From~\eqref{eqn:EEequalsuperposition}, we finally arrive at
\begin{equation}
   \boxed{ S_A=(Area)\log N-2\log N.}\label{main}
\end{equation}
Here, $(Area)$ is the number of vertex operators that have support on both of $A$ and its complement, B, which can be regarded as the number of clock states with type 1 (i.e., clock states located at vertices) surrounding the disk geometry $A$ (see Fig.~\ref{area}). In the present case, it is given by $(Area)=4l_y+2(l_x-2)$.
One can check that the entanglement entropy of the disk geometry for the generic ground state~\eqref{generic} gives the same answer as~\eqref{main}.

The first term in~\eqref{main} is nothing but the area law term and the second term, which is the sub-leading term of the entanglement entropy, is of particular importance. It is ``topological"; one can confirm that the second term remains the constant when we deform the shape of the disk geometry retaining its topology.\footnote{One can verify that 
the sub-leading order term of~\eqref{main} is topological by the prescription given in~\cite{LevinWen2006TEE,preskillkitaev2006} where the topological entanglement entropy is extracted from a linear combination of the entanglement entropy of disk geometries in a way that the area law terms are suppressed.} Therefore, we have found the same scaling behavior of the entanglement entropy as the one obtained in conventional topologically ordered phases~\eqref{ee}. However, there is a crucial deference between the two. In the case of the conventional topologically ordered phases, the topological entanglement entropy $\gamma$ is related to the total quantum dimension via $\gamma=\log\sqrt{\sum_ad^2_a}$, where $a$ labels the distinct types of fractional excitations. On the contrary, in our case, such a relation does not hold as the number of distinct number of fractional excitations varies depending on the system size [\eqref{qd}] and the topological entanglement entropy takes a constant number, $\gamma=-2\log N$ irrespective to the system size, implying the incapability to associate $\gamma$ to the total quantum dimension.
In the next section, we give a physical interpretation of this result in the simplest setting, i.e., $N=2$.

\begin{table}[h]
  \begin{tabular}{c|c|c} 
  &$\ket{\xi_{00,00}}$ & generic ground state $\ket{\zeta}$~\eqref{generic}\\\hline
single row I [Fig.~\ref{fig:singlerow}a]&$n_x\log N-\log \Gamma$&$n_x\log N-\log \Gamma+\mu_1$~\eqref{r1prime}\\
single row II [Fig.~\ref{fig:singlerow}b]&$n_x\log N$&$n_x\log N+\mu_2$~\eqref{nu1}~\eqref{nu2}\\
single column [Fig.~\ref{fig:singlerow}c]&$2n_y\log N-\log N$&$2n_y\log N-\log N+\mu_3$~\eqref{nu3}\eqref{nu4}\\
disk geometry [Fig.~\ref{disk}]&$(Area)\log N-2\log N$&$(Area)\log N-2\log N$
  \end{tabular}
  \caption{Summary of the entanglement entropy $S_A$ of various subsystem geometries with respect to the ground state~$\ket{\xi_{00,00}}$ and the one with the generic ground state~\eqref{generic}. Here, $\Gamma=N\times\gcd(N,n_x)$ and (\textit{Area}) represents the number of clock states with type 1 surrounding the disk geometry.}\label{table1}
\end{table}
We summarize the results obtained in this section in Table.~\ref{table1}

\section{Alternative interpretation of the result in the simplest case}\label{sec4}

In this section, we give an intuitive understanding of the topological entanglement entropy found in the second term of~\eqref{main} by focusing on the simplest case of $N$, i.e., $N=2$. To this end, discussion presented in Sec.~\ref{sec2b} is of usefulness.

Recalling the 
argument in Sec.~\ref{sec2b}, in the case of $N=2$,
depending on whether the length of the lattice in the $x$-direction is even or odd, the GSD, accordingly, the total quantum dimension changes whereas the topological entanglement entropy gives $-2\log2$. Even though we have the topological entanglement entropy $-2\log2$ irrespective to the length of the lattice in the $x$-direction being even or odd, its topological origin to have such a number is rather different in these two cases.  \par
\begin{figure}
    \begin{center}
       \begin{subfigure}[h]{0.50\textwidth}
  \includegraphics[width=\textwidth]{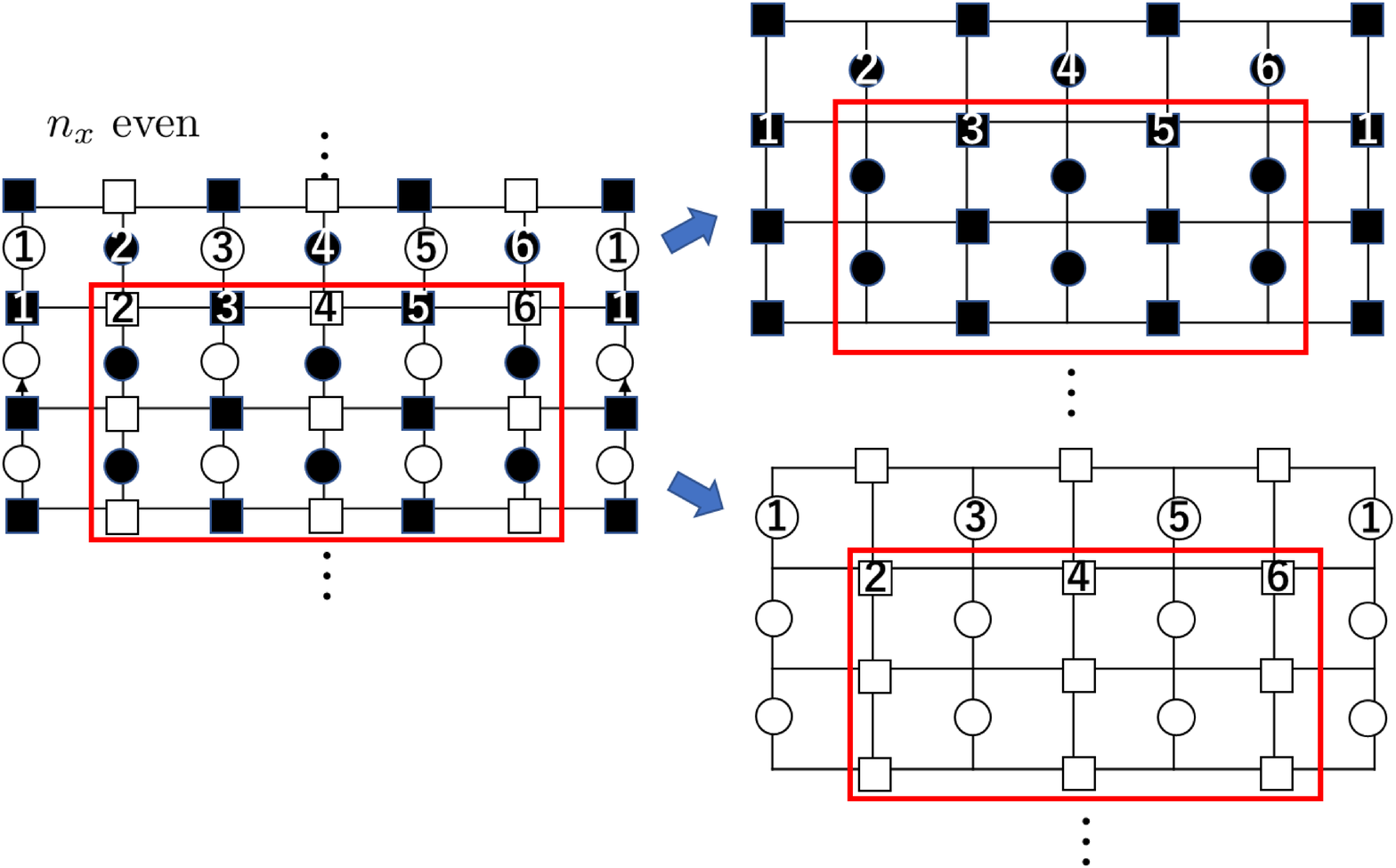}
         \caption{}\label{even}
             \end{subfigure}
             \hspace{5mm}
               \begin{subfigure}[h]{0.50\textwidth}
    \includegraphics[width=\textwidth]{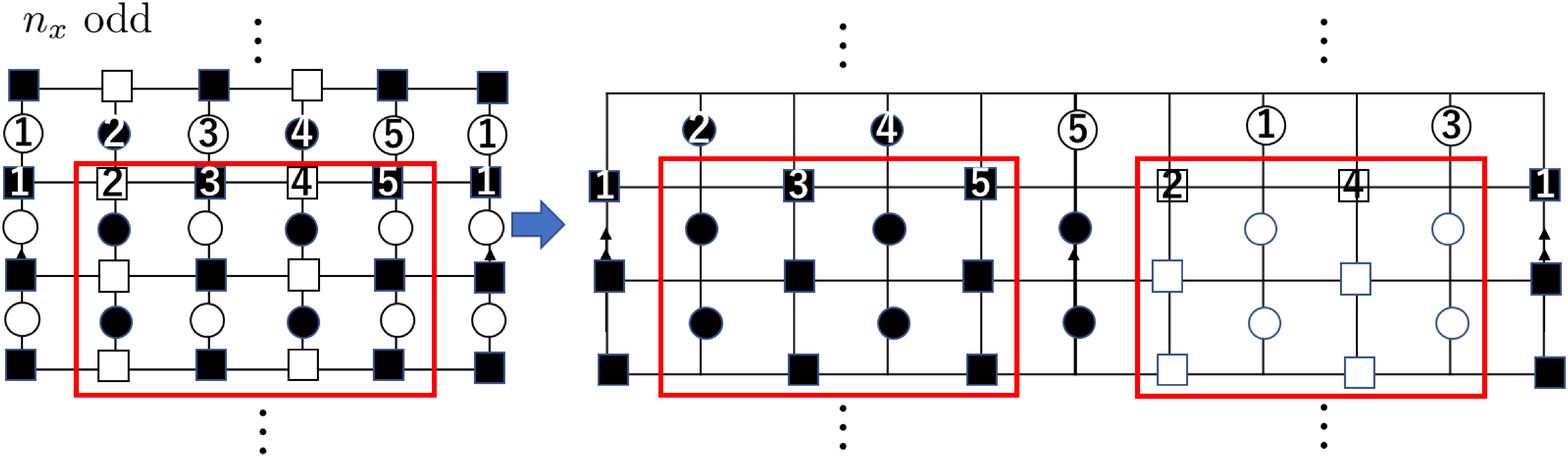}
         \caption{}\label{odd}
             \end{subfigure} 
 \end{center}
 \caption{(a) In the case of $n_x$ being even. One can decompose the lattice into two. The periodic boundary condition is imposed so that the left and right edge are identified.
 (b) In the case of $n_x$ being odd. We double the lattice by going around the torus twice, and omit vertices and links as explained in the main text. 
 The red frame corresponds to the subsystem $A$ with disk shape considered in the previous section.   }
 \end{figure}

As seen from~\eqref{ppv2}, in the case of $N=2$ and $n_x$ being even, 
the mutually commuting terms of Hamiltonian get simplified, allowing us to decompose the Hamiltonian into two sectors, each of which describes the~$\mathbb{Z}_2$ toric code. 
Such decomposition is shown in Fig.~\ref{even} where we separate vertices with even number $x$ coordinate (white squares) and the ones with odd number (black squares) as well as vertical links with odd number $x$ coordinate (white dots) and the ones with even number~(black dots). 
Correspondingly, the entanglement entropy $S_A$ is decomposed into two, each of which is the one of the disk geometry in the $\mathbb{Z}_2$ toric code. Since the topological entanglement entropy of the disk geometry in the $\mathbb{Z}_2$ toric code is known to be $-\log2$~\cite{KitaevPreskill2006TEE,LevinWen2006TEE}, in total, the topological entanglement entropy gives $2\times(-\log2)=-2\log2$. \par
On the contrary, in the case of $n_x$ being odd, such decomposition~\eqref{group} is not valid, thus the Hamiltonian describes one $\mathbb{Z}_2$ toric code. To see this more clearly, 
recalling the fact that each commuting term of the Hamiltonian involves next nearest neighboring Pauli operators in the $x$-direction, 
we think of doubling the lattice, obtained by going around the torus in the $x$-direction twice and rearranging the vertices and links. 
\par
To be more specific to what we have just mentioned, consider a geometry portrayed in Fig.~\ref{odd}. We double the lattice obtained by traveling around the torus in the $x$-direction twice. 
In the first half of the doubled lattice, we omit the vertices 
with even number $x$ coordinate and vertical links with odd number $x$ coordinate whereas in the latter half, we discard the vertices with with odd number $x$ coordinate and vertical links with even number $x$ coordinate. 
Recalling~\eqref{ppa}, one is convinced that the Hamiltonian describes one $\mathbb{Z}_2$ toric code on the torus. By this rearrangement, the entanglement entropy~$S_A$ amounts to be the one of two spatially separated disks in the~$\mathbb{Z}_2$ toric code. The topological entanglement entropy is found to be $-2\log2$ -- the same value as the case of $n_x$ even.


\section{Generic lattices}\label{sec7}
As we have seen in the previous argument, our model contains the second order spatial derivative, involving the next nearest neighboring sites. Such a property can be straightforwardly described by the Laplacian, the graph theoretical analogue of the second order spatial derivatives with which one can evaluate the entanglement entropy. 
 In this section, 
 we further explore the entanglement entropy of the model on \textit{generic lattices},  rather than standard square one to see the interplay between quantum entanglement and geometry of the system.
\subsection{Laplacian and Hamiltonian}
To this end, 
let us start with reviewing a few terminologies in the graph theory. A \textit{graph} $G=(V,E)$ which is a pair
consisting of a set of vertices~$V$ and a set of edges $E$ comprised of pairs of vertices~$\{v_i,v_j\}$. In the rest of the work, we assume that the graph is \textit{connected}, meaning there is a path from a vertex to any other vertex, and that the graph does not have an edge that emanates from and terminates at the same vertex.
We also introduce 
two quantities, deg$(v_i)$ and $l_{ij}$. The former one,~deg$(v_i)$ denotes the \textit{degree} of the vertex $v_i$, i.e., the number
of edges emanating from the vertex $v_i$ and the latter one, $l_{ij}$ represents the number of edges between two vertices $v_i$ and $v_j$~(We have $l_{ij}=0$ when there is no edge between two vertices, $v_i$ and $v_j$.).
Using these two quantities, 
\textit{Laplacian matrix} of the graph, which is the analogue of the second order derivative operator $\partial_x^2$ on a graph, is defined.
For a given graph $G=(V,E)$, the Laplacian matrix $L$ (which we abbreviate as Laplacian in the rest of this work) is the matrix with rows and columns indexed by the elements of vertices~$\{v_i\}\in V$, with
\begin{equation}
    L_{ij}=\begin{cases}\text{deg}(v_i)\;(i=j)
    \\-l_{ij}\;(i\neq j)
    \end{cases}.\label{laplacian}
\end{equation}
The Laplacian is singular 
due to the connectivity of the graph. (Summing over all rows or columns gives zero.)
As an example, the Laplacian of the cycle graph $C_4$ (i.e., a square) consisting of four vertices and four edges, where there is a single edge between a pair of vertices, is given by
\begin{equation*}
    L=\begin{pmatrix}
2 & -1&&-1 \\
-1 & 2&-1 &\\
&-1&2&-1\\
-1&&-1&2
\end{pmatrix}.
\end{equation*}
  \begin{figure}
    \begin{center}
         
       \includegraphics[width=0.84\textwidth]{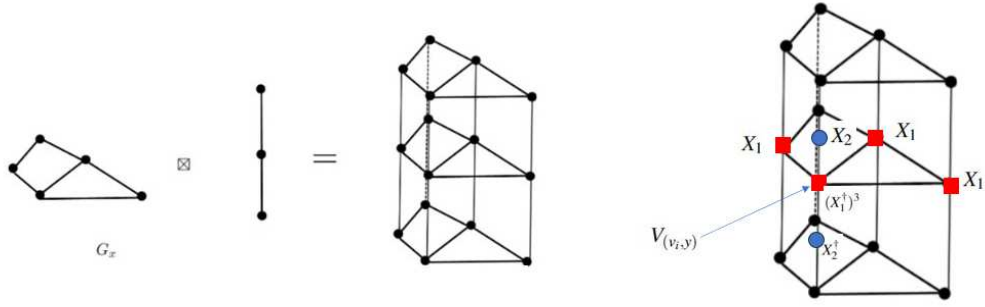}
        \end{center}
       
      \caption{Left: One example of the 2D lattice consisting of a connected graph $G_x$ and the 1D line. Right: One of the term, $V_{(v_i,y)}$ defined in~\eqref{pp02} is shown in the same example of the lattice.
 }
        \label{latticemodel}
   \end{figure}
   It is known that by introducing invertible matrices over integer $P$, $Q$ corresponding to linear operations on rows and columns of the Laplacian, respectively, the Laplacian of any connected graph can be transformed into a diagonal form (\textit{Smith normal form}):\footnote{The last entry is zero, mirroring the fact that the graph is connected.}
\begin{equation}
    PLQ=\text{diag}(u_1,u_2,\cdots,u_{n-1},0). \label{snf0}
\end{equation}
The diagonal entries $u_i$ are referred to as the \textit{invariant factors}.\par
With these terminologies, we introduce generic 2D lattice on which the higher rank topological phase is placed.
We consider 2D lattices constructed by the product of the graph $G(V,E)$ and 1D line. An example of such a lattice is shown in Fig.~\ref{latticemodel} left. The lattice is constructed in such a way that copies of the graphs are stack along the vertical direction. \footnote{Note that the lattice obtained this way is not embeddable in flat 2D space. Rather, it is defined on an abstract 2D cell complex. Indeed, the graph
consists of vertices and edges, corresponding to $0$- and $1$-simplices. Due to this reason, we regard our lattice as 2D. }

We introduce two 
types of the generalized $N$-qubit states ($\mathbb{Z}_N$ clock states) on this 2D lattice. 
The first type of the clock states are located on the vertices of the graph in the horizontal direction, whereas the ones with the second type on the vertical links. We denote the coordinate of the first clock states by~$(v_i,y)$ where $v_i$ represents a vertex of the graph and $y$ does the height taking integer values in the unit of lattice spacing. Analogously, the coordinate of the second clock states are denoted by $(v_i,y+\frac{1}{2})$, where the second element corresponds to the edge between vertices located at~$(v_i,y)$ and $(v_i,y+1)$.\par
Having defined the 2D lattice, we introduce the Hamiltonian which is generalization of the one studied in the previous sections to an arbitrary graph. 
We represent the local Hilbert space of the two types of the clock states as $\ket{a}_{(v_i,y)}$ and $\ket{b}_{(v_i,y+1/2)}$ ($a,b\in\mathbb{Z}_N$). Further, define operators $X_{1,(v_i,y)}$, $Z_{1,(v_i,y)}$, and  $X_{2,(v_i,y+1/2)}$, $Z_{2,(v_i,y+1/2)}$ that act on these states as
\begin{eqnarray}
Z_{1,(v_i,y)}\ket{a}_{(v_i,y)}=\omega^a\ket{a}_{(v_i,y)},\;Z_{2,(v_i,y+1/2)}\ket{b}_{(v_i,y+1/2)}=\omega^b\ket{b}_{(v_i,y+1/2)}\nonumber\\
X_{1,(v_i,y)}\ket{a}_{(v_i,y)}=\ket{a+1}_{(v_i,y)},\;X_{2,(v_i,y+1/2)}\ket{b}_{(v_i,y+1/2)}=\ket{b+1}_{(v_i,y+1/2)}
\end{eqnarray}
with
$\omega=e^{2\pi i/N}$.
With these notations, 
we define following two types of operators that we dub vertex and plaquette operators at each vertex and vertical link:
\begin{eqnarray}
V_{(v_i,y)}\vcentcolon=X_{2,(v_i,y+1/2)}X_{2,(v_i,y-1/2)}^{\dagger}(X_{1,(v_i,y)}^{\dagger})^{\text{deg}(v_i)}\prod_j(X_{1,(v_j,y)})^{l_{ij}},\nonumber\\
P_{(v_i,y+1/2)}\vcentcolon=Z_{1,(v_i,y+1)}^{\dagger}Z_{1,(v_i,y)}Z_{2,(v_i,y+1/2)}^{\text{deg}(v_i)}\prod_j(Z_{2,(v_j,y+1/2)}^{\dagger})^{l_{ij}}.\label{pp02}
\end{eqnarray}
Here, deg$(v_i)$ 
is the number of edges in the $x$-direction which emanate from the vertex $v_i$ whereas $l_{ij}$ gives the number of edges in the $x$-direction between two vertices with coordinate~$(v_i,y)$ and $(v_j,y)$, in accordance with the matrix element of the Laplacian~\eqref{laplacian}. It is straightforward to check terms given in~\eqref{pp02} commute with each other. 
We demonstrate one example of the term in the right Fig.~\ref{latticemodel}.
 Hamiltonian is defined by
\begin{equation}
    H=-\sum_{(v_i,y)}V_{(v_i,y)}-\sum_{(v_i,y+1/2)}P_{(v_i,y+1/2)}\label{hamiltonian2}+h.c.
\end{equation}
When we impose the periodic boundary condition in the $y$-direction, 
by the same reasoning presented in Sec.~\ref{gsd}, one finds that the GSD is given by~\cite{ebisu2209anisotropic}
\begin{equation}
   {\text{GSD}=\bigl[N\times\text{gcd}(N,u_1)\times\cdots \times\text{gcd}(N,u_{n-1})\bigr]^2.}
\end{equation}
Here, positive integers, $u_i$ denote the invariant factors of the Laplacian~\eqref{snf0}.


 \subsection{Entanglement entropy}
\begin{figure}
    \begin{center}
        \begin{subfigure}[h]{0.12\textwidth}
    \includegraphics[width=\textwidth]{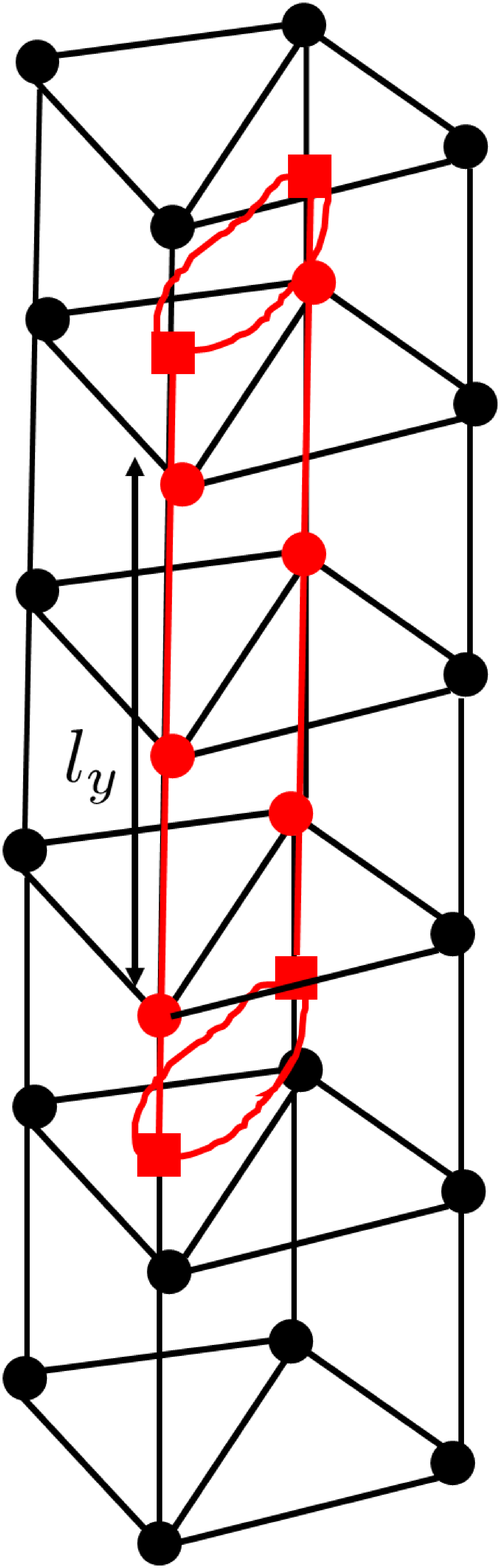}
         \caption{}\label{tower}
             \end{subfigure} 
               \begin{subfigure}[h]{0.42\textwidth}
    \includegraphics[width=0.75
    \textwidth]{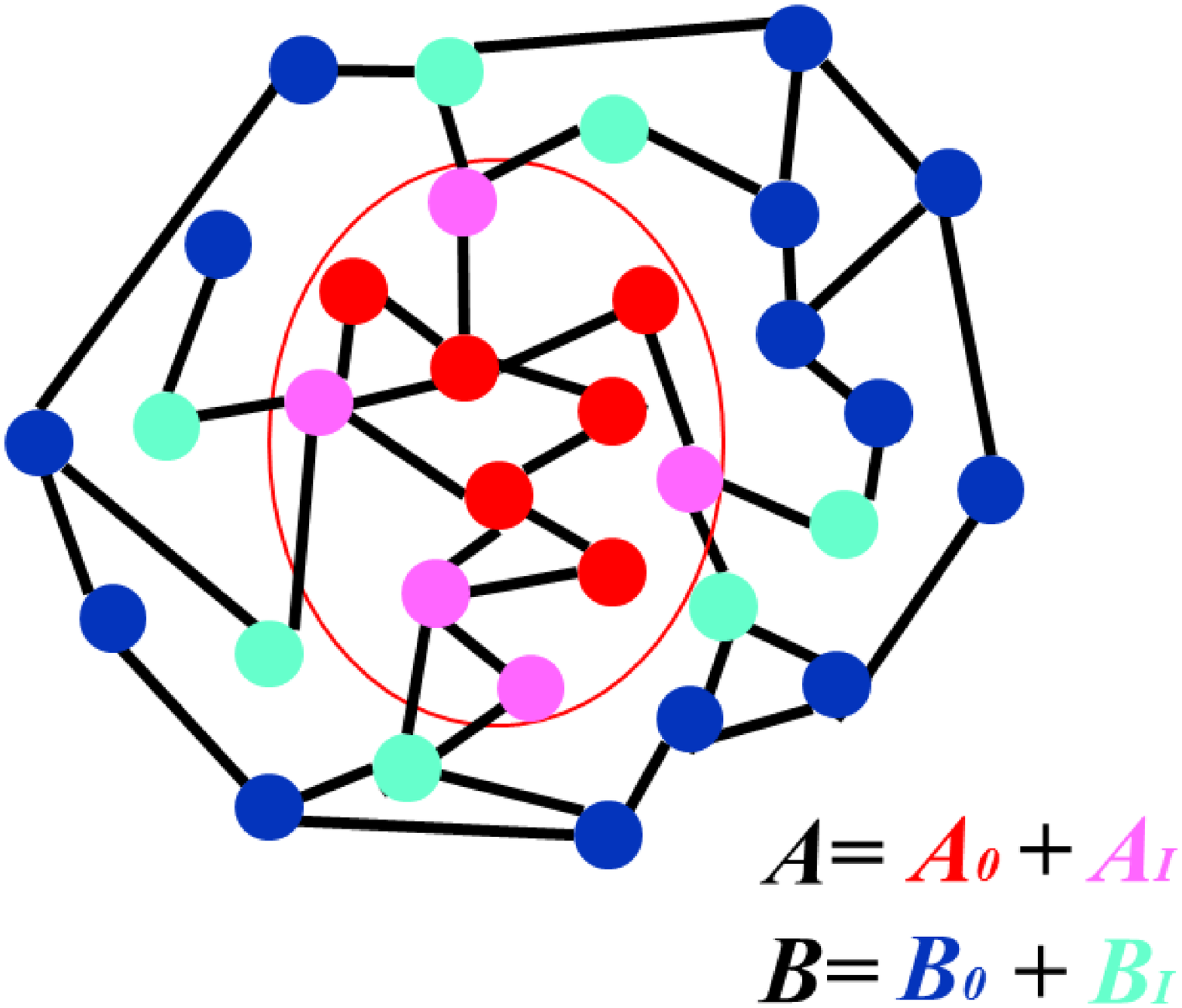}
         \caption{}\label{config}
             \end{subfigure} 
                    
             \caption{(a) Example of the subsystem A in a shape of a cylinder geometry consisting of the sub-graph and the 1D line. The clock states belonging to the subsystem A are marked by red colors. (b)~Classification of the vertices of the graph at $(y_0\leq y\leq y_0-1+l_y)$. Inside the subsystem A, the vertices which are (are not) connected with the ones outside A are represented by $A_I$ ($A_0$), marked by pink (red) color. Also, outside A, the vertices which are (are not) connected with the ones in A are denoted by $B_I$ ($B_0$), marked by light blue (blue).  }
 \end{center}
 \end{figure}

Now we are in a good place to study the entanglement entropy of our model on a graph.     
We set the sub-system A as the ``cylinder geometry" consisting of
the sub-graph of $G$ and the vertical line with length $l_y$, i.e., the cylinder encloses $l_y$ copies of the sub-graphs.
The clocks states belonging to the subsystem A are the ones located on the vertices of the sub-graph within the cylinder and the ones on the vertical links inside the cylinder and vertical links that cross the top and bottom face of the cylinder. We also denote the compliment of the subsystem A as B.
For simplicity, we set the sub-graph to be connected. An example is shown in Fig~\ref{tower}. We set the $y$ coordinate of vertices inside the subsystem A to be $(y_0\leq y\leq y_0-1+l_y)$.
\par
 For later convenience, 
at each $y$, $(y_0\leq y\leq y_0-1+l_y)$, we label vertices of the graph classified by following four groups:
\begin{eqnarray}
    A_0&=&\{\text{vertices inside A which are not connected with the ones outside A}\}\nonumber\\
        A_I&=&\{\text{vertices inside A which are connected with the ones outside A}\}\nonumber\\
         B_I&=&\{\text{vertices outside A which are connected with the ones inside A}\}\nonumber\\
        B_0&=&\{\text{vertices outside A which are not connected with the ones inside A}\}\label{class}
\end{eqnarray}
An example is demonstrated in Fig.~\ref{config}. We denote the number of such vertices by $n_{A_0}$, $n_{A_I}$, $n_{B_I}$, and $n_{B_0}$.
In accordance with the vertices of the four classes~\eqref{class}, we rearrange the order of the vertices so that vertices of $A_0$ come first, followed by the ones of $A_I$, $B_I$, and $B_0$. In this order of the vertices, 
we write the Laplacian of the graph in the following form
\begin{equation}
    L=\begin{pmatrix}
L_{A_0A_0} &L_{A_0A_I}&& \\
L_{A_IA_0} & L_{A_IA_I}&L_{A_IB_I}& \\
&L_{B_IA_I}&L_{B_IB_I}&L_{B_IB_0}\\
&&L_{B_0B_I}&L_{B_0B_0}
\end{pmatrix},\label{subm}
\end{equation}
where $L_{**}$ describes the sub-matrix of the Laplacian. For instance, the matrix $L_{A_0A_0}$ is the $n_{A_0}\times n_{A_0}$ matrix indexed by vertices which belong to $A_0$. After preparing terminologies, we are going to show the following:
\begin{frm-thm}[Entanglement entropy of a sub-graph]
Consider the stabilizer model on the 2D lattice~\eqref{hamiltonian2} obtained by the product of the graph and 1D line with periodic boundary condition in the vertical direction. The entanglement entropy of the subsystem A with cylinder geometry, constructed by a sub-graph and the 1D line with length $l_y$ with respect to the ground state~$\ket{\psi}$
is given by
\begin{eqnarray}
    S_A=
  \biggl[(Area)\log N-2l_y\log(\prod_i\gcd\bigl(N,r_i)\bigr)-l_y|n_{A_I}-n_{B_I}|\log N\biggr]\nonumber\\
    -\log\biggl(\frac{N^{\min(n_{A_I},n_{B_I})}}{\prod_i\gcd\bigl(N,r_i)}\biggr)-2\log\bigl(\prod_j\gcd(N,s_j)\bigr).
    \label{main2}
\end{eqnarray}
Here, $(Area)$ represents the number of the vertex operators that have support on both of A and B, which is associated with the number of the clock states located around the border of A and B. Also, 
the number $r_i$ and $s_j$ represents invariant factors of the sub-matrix of the Laplacian via (SNF represents the Smith normal form)
\begin{equation}
    L_{A_IB_I}\xrightarrow[]{SNF}  \begin{pmatrix}
    r_1&&\\
    &\ddots&\\
    &&r_{p}\\
    &&\\
    &&\\
    &&
    \end{pmatrix},
\begin{pmatrix}
   L_{A_0A_0} &L_{A_0A_I}&\\
L_{A_IA_0} & L_{A_IA_I}&L_{A_IB_I}
    \end{pmatrix} \xrightarrow[]{SNF} \begin{pmatrix}
    s_1&&&&\\
    &\ddots&&&\\
    &&s_{n_{A_0}+n_{A_I}}&&
     \end{pmatrix}.
\end{equation}
\label{th}
\end{frm-thm}
\textit{Proof}.

\begin{figure}
    \begin{center}
         
       \includegraphics[width=0.10\textwidth]{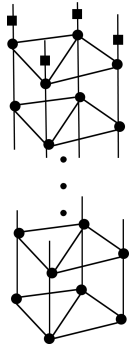}

 \end{center}
       
      \caption{We accommodate boundary condition on the lattice so that in the bottom the lattice terminates the graph whereas on the top, the lattice terminates with the vertical links. One can verify that the there is no constraint on the stabilizes on entire lattice, simplifying the problem.
 }
        \label{toweer2}
   \end{figure}

In order to find the entanglement entropy, we resort to the formula~\eqref{eqn:EEequalsuperposition}.
Generically, it is cumbersome to evaluate $|G_A|$ and $|G_B|$ due to the constraint on the stabilizers. As we have seen in Sec.~\ref{sec3}, 
the complication stems from the fact that there is non-trivial GSD when we impose the periodic boundary condition on the lattice, giving rise to constrains on the multiplication of the stabilizers on entire lattice. 
While it is still possible to implement the similar approach as the one discussed in Sec.~\ref{sec3} to identify the entanglement entropy, here we employ an alternative simpler way to obtain the entanglement entropy.
We accommodate an appropriate boundary condition on the system in the vertical direction, instead of the periodic boundary condition. Consider the 2D lattice that we introduced in~\eqref{hamiltonian2} with finite length in the vertical direction so the bottom boundary terminates with a graph $G$ and the top boundary ends with vertical links (see Fig.~\ref{toweer2}). With this boundary condition, one can check that the GSD is trivial and there is no constraint on the multiplication of stabilizers on entire lattice, simplifying the problem.
The boundary that we consider here is intriguing on its own right. It is reminiscent of smooth and rough boundary 
of the toric code, where anyons with magnetic and electric charge are condensed~\cite{bravyi1998quantum}.\par

After accommodating the boundary, now we turn to calculation of the entanglement entropy based on~\eqref{eqn:EEequalsuperposition} assuming $n_{B_I}\leq n_{A_I}$ (The proof in the case of $n_{B_I}> n_{A_I}$ is similarly discussed.).
  From~\eqref{eqn:EEequalsuperposition}, one needs to evaluate the number of product of vertex operators that act exclusively on A or B. 
We denote the number of \textit{individual} vertex terms 
   that act within A and B as $|\tilde{G}_A|$, and $|\tilde{G}_B|$. To find 
   $|G_A|$ and $|G_B|$, we also need to
 think of the multiplication of the vertex terms which act exclusively on A and B. Representing these numbers as $\Gamma_A$ and $\Gamma_B$, $|G_A|$ and $|G_B|$ is written as \begin{equation}
     |G_A|=|\tilde{G}_A|\times \Gamma_A,\;\;|G_B|=|\tilde{G}_B|\times \Gamma_B.\label{nini}
 \end{equation}
 Below we identify $\Gamma_A$ and $\Gamma_B$.

   At given $y$ $(y_0\leq y\leq y_0-1+l_y)$, we consider the following product of the vertex terms:
   \begin{equation}
      \prod_{v_i\in A_I} V_{(v_i,y)}^{\alpha_i}\;(\alpha_i\in\mathbb{Z}_N),\label{pp1}
   \end{equation}
   which can be rewritten as 
   \begin{equation}
        \prod_{v_i\in A_I} V_{(v_i,y)}^{\alpha_i}=\biggl[\prod_{i\in A_I}X_{2,(v_i,y+1/2)}^{-\alpha_i}X_{2,(v_i,y-1/2)}^{\alpha_i}\biggl]\times \prod_{j\in A_0+A_I+B_I}X_{1,(v_j,y)}^{\beta_j}\;(\beta_j\in\mathbb{Z}_N).\label{hey}
   \end{equation}
   Defining a $n_{A_I}$- and $n_{A_0+A_I+B_I}$-dimensional vector $\bm{\alpha}$ and $\bm{\beta}$ whose element is given by $\alpha_i$ and $\beta_j$ respectively, and referring to~\eqref{subm}, it follows that the vector $\bm{\alpha}$ and $\bm{\beta}$ are related via sub-matrix of the Laplacian as
\begin{equation}
   \bm{\beta}=\begin{pmatrix}
       L_{A_0A_I}\\
       L_{A_IA_I}\\
       L_{B_IA_I}
   \end{pmatrix}\bm{\alpha}.
\end{equation}
Suppose the last $n_{B_I}$ entries of $\bm{\beta}$ are zero. Then the product~\eqref{hey} is what we want: multiplication of the vertex operators that act exclusively on $A_I$, contributing to $\Gamma_A$. Thus, the number of such product amounts to be the number of solution of
\begin{equation}
      L_{B_IA_I}\bm{\alpha}=\bm{0}\mod N.\label{con1}
\end{equation}
Introducing $n_{B_I}\times n_{B_I}$ and $n_{A_I}\times n_{A_I}$ invertible integer matrix, $U$ and $V$, one can transform the $n_{B_I}\times n_{A_I}$~matrix $L_{B_IA_I}$ into the Smith normal form as
\begin{equation}
    UL_{B_IA_I}V=  \begin{pmatrix}
    r_1&&\\
    &\ddots&\\
    &&r_{n_{A_I}}\\
    &&\\
    &&\\
    &&
    \end{pmatrix}.\label{snf1}
\end{equation}
From this form~\eqref{snf1}, we have 
\begin{equation}
     L_{B_IA_I}\bm{\alpha}=\bm{0}\Leftrightarrow  \begin{pmatrix}
    r_1&&\\
    &\ddots&\\
    &&r_{n_{A_I}}\\
    &&\\
    &&\\
    &&
    \end{pmatrix}\tilde{\bm{\alpha}}=\bm{0}\mod N,
\end{equation}
where $\tilde{\bm{\alpha}}\vcentcolon=V^{-1}\bm{\alpha}$.
The $i$th element of $\tilde{\bm{\alpha}}$ is subject to
\begin{align}
    r_i\tilde{\alpha}_i=0\mod N,
\end{align}
from which it follows that $r_i$ takes $\gcd(N,r_i)$ distinct values, where gcd stands for the greatest common divisor. 
Therefore, the number of the solution of~\eqref{con1} is given by
\begin{equation*}
    \prod_{i=1}^{n_{A_I}}\gcd(N,r_i).\label{sol55}
\end{equation*}
Since we have considered the product of the vertex operators defined on $A_I$ that act trivially on $B_I$ at given $y$ $(y_0\leq y\leq y_0-1+l_y)$, there are 
in total
\begin{equation}
    \biggl[\prod_{i=1}^{n_{A_I}}\gcd(N,r_i)\biggr]^{l_y}\;(\vcentcolon=\Gamma_1)\label{gamma1}
\end{equation}
of the product of the vertex operators defined on $A_I$ which acts exclusively on $A_I$, contributing to $\Gamma_A$.\par
We can analogously discuss the number of product of the vertex operators located on $B_I$ which act trivially on $A_I$, contributing to $\Gamma_B$. Suppose the following product at given $y$ $(y_0\leq y\leq y_0-1+l_y)$ 
\begin{equation}
    \prod_{j\in B_I}V_{(v_j,y)}^{\eta_j}\;(\eta_{j}\in\mathbb{Z}_N)
\end{equation}
acts trivially on $A_I$. By the similar augment presented around~\eqref{con1}, the number of such product amounts to be the number of solution of
\begin{equation}
    L_{A_IB_I}\bm{\eta}=\bm{0}\mod N,\label{con2}
\end{equation}
where $\bm{\eta}$ denotes $n_{B_I}$-dimensional vector whose entry is given by $\eta_j$. Similarly to~\eqref{snf1}, we transform the matrix $L_{A_IB_I}$ into the Smith normal form.
Introducing $n_{A_I}\times n_{A_I}$ and $n_{B_I}\times n_{B_I}$ invertible integer matrix, $U^\prime$ and $V^\prime$, the Smith normal form reads
\begin{equation}
   U^\prime L_{A_IB_I}V^\prime=  \left(
\begin{array}{ccc|cccc}
r_1&&&&0&&\\
    &\ddots&&&&\ddots&\\
    &&r_{n_{A_I}}&&&&0
\end{array}
\right)
.\label{snf2}
\end{equation}
Since $L_{B_IA_I}^T=L_{A_IB_I}$, the smith normal form of the $n_{A_I}\times n_{B_I}$ matrix  
$ L_{A_IB_I}$ contains the same diagonal entries as~\eqref{snf1}. Eq.~\eqref{con2} is rewritten as
\begin{eqnarray}
     \left(
\begin{array}{ccc|cccc}
 r_1&&&&0&&\\
    &\ddots&&&&\ddots&\\
    &&r_{n_{A_I}}&&&&0
\end{array}
\right)\tilde{\bm{\eta}}=\bm{0}\mod N
\end{eqnarray}
with $\tilde{\bm{\eta}}\vcentcolon=V^{\prime-1}\bm{\eta}$. The first  $n_{A_I}$ entries of  $\tilde{\bm{\eta}}$ are subject to $\tilde{\eta}_i=r_i\;(1\leq i\leq n_{A_I})$ whereas there is no constraint on the last $n_{B_I}-n_{\Ai}$ entries of $\tilde{\bm{\eta}}$. Hence, the number of solution satisfying~\eqref{con2} is given by 
\begin{equation}
    \prod_{i=1}^{n_{A_I}}\gcd(N,r_i)\times N^{(n_{B_I}-n_{A_{I}})}.\label{sol}
\end{equation}
Taking into the $y$-direction, we have
\begin{equation}
    \biggl[\prod_{i=1}^{n_{A_I}}\gcd(N,r_i)\times N^{(N_{B_I}-n_{A_{I}})}\biggr]^{l_y}\;(\vcentcolon=\Gamma_2)\label{gamma2}
\end{equation}
product of the vertex operators defined on $B_I$
   which act trivially on $A_I$, contributing to $\Gamma_B$.

\begin{figure}
    \begin{center}
         
       \includegraphics[width=0.24\textwidth]{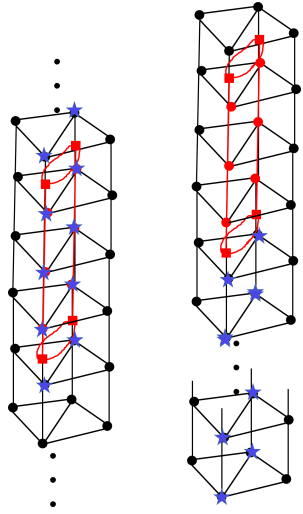}

 \end{center}
       
      \caption{Left: Multiplication of the stabilizes $V_{v_i,y}$ with blue stars gives rise to the operators $X_1$'s
and $X_2$'s  that act exclusively on B. Right: Multiplication of the stabilizes $V_{v_i,y}$ with blue stars gives rise to the operators $X_1$'s
and $X_2$'s  that act exclusively on A. The clock states with red color belong to the subsystem~A.}
        \label{tower3}
   \end{figure}

In addition to~\eqref{gamma1}~\eqref{gamma2}, there are product of the vertex operators along both of horizontal and vertical directions, which contributes to $\Gamma_A$ and $\Gamma_B$. At given $y$, let us consider the following product
\begin{equation}
    \prod_{i\in A_0+A_I}V_{(v_i,y)}^{\delta_i}\;(\delta_i\in\mathbb{Z}_N).\label{hi2}
\end{equation}
Suppose~\eqref{hi2} does not have $X_1$ operators in the horizontal direction. Recalling the previous argument presented around~\eqref{con1} and~\eqref{con2}, such a condition is described by
\begin{equation}
    \begin{pmatrix}
        L_{A_0A_0}&L_{A_0A_I}\\
        L_{A_IA_0}&L_{\Ai\Ai}\\
        &L_{\Bo\Ai}
        \end{pmatrix}
       \bm{\delta} =\bm{0}
\mod N    \label{mp}
\end{equation}
By transforming the matrix on the left hand side of~\eqref{mp} into the Smith normal form as
\begin{equation}
      \begin{pmatrix}
        L_{A_0A_0}&L_{A_0A_I}\\
        L_{A_IA_0}&L_{\Ai\Ai}\\
        &L_{\Bo\Ai}
        \end{pmatrix}\to\begin{pmatrix}
            s_1&&\\
            &\ddots&\\
            &&s_{n_{\Ai+\Ao}}\\
            &&\\
            &&
        \end{pmatrix},
\end{equation}
there are
\begin{equation}
\Gamma_3\vcentcolon=\prod_{i=1}^{n_{\Ai+\Ai}}\gcd(N,s_i)\label{sol2}
\end{equation}
solutions. Multiplying~\eqref{hi2} along vertical direction give rise to stabilisers, which contribute to $\Gamma_A$ and $\Gamma_B$. Indeed, 
\begin{equation}
    \prod_{y=y_0-1}^{y_0+l_y}\bigl[  \prod_{i\in A_0+A_I}V_{(v_i,y)}^{\delta_i}\bigr]
\end{equation}
has support only on $B$, which contributes to $\Gamma_B$ (Fig.~\ref{tower3} left). Likewise, the product
\begin{equation}
     \prod_{y=1}^{y_0}\bigl[  \prod_{i\in A_0+A_I}V_{(v_i,y)}^{\delta_i}\bigr]
\end{equation}
acts exclusively on $A$, contributing to $\Gamma_A$ (Fig.~\ref{tower3} right).
\par
We also consider the product of the vertex operators defined on $\Ao+\Ai+\Bi$:
\begin{equation}
    \prod_{i\in\Ao+\Ai+\Bi}V_{(v_i,y)}^{\sigma_i}\label{sigma}
\end{equation}
Assuming it does not have $X_1$ operators on $\Ao+\Ai$. Such a condition can be described by
\begin{equation}
\begin{pmatrix}
      L_{A_0A_0} &L_{A_0A_I}&\\
L_{A_IA_0} & L_{A_IA_I}&L_{A_IB_I}
    \end{pmatrix}\bm{\sigma}=\bm{0}\mod N.\label{sol3}
\end{equation}
Multiplying~\eqref{sigma} along $y$
\begin{equation}
    \prod_{y=y_0}^{y_0-1+l_y}\prod_{i\in\Ao+\Ai+\Bi}V_{(v_i,y)}^{\sigma_i}\label{prods}
\end{equation}
yields the stabilizer that acts within B, contributing to $\Gamma_B$. Thus, we need to evaluate the number of solutions satisfying~\eqref{sol3}. However, some of solutions are redundant to what we have already identified previously~\eqref{sol}\eqref{sol2}. Therefore, the number of product in question which contributes to $\Gamma_B$~\eqref{prods} is equivalent to 
the number of solutions of~\eqref{sol3} divided by the numbers given in~\eqref{sol}\eqref{sol2}:
\begin{equation}
    \frac{\text{The number of solutions of~\eqref{sol3}}}{\prod_{i=1}^{n_{A_I}}\gcd(N,r_i)\times N^{(N_{\Bi}-N_{\Ai})}\times\prod_{j=1}^{n_{\Ai+n_{A_0}}}\gcd(N,s_j)}.\label{42}
\end{equation}
We transform the matrix in~\eqref{sol2} into the Smith normal form as
\begin{equation}
    \begin{pmatrix}
   L_{A_0A_0} &L_{A_0A_I}&\\
L_{A_IA_0} & L_{A_IA_I}&L_{A_IB_I}
    \end{pmatrix} \xrightarrow[]{SNF} 
     \left(
\begin{array}{ccc|cccc}
 s_1&&&&0&&\\
    &\ddots&&&&\ddots&\\
    &&s_{n_{A_0}+n_{A_I}}&&&&0
\end{array}
\right),
\end{equation}
from which it follows that the the denominator of~\eqref{42} is given by $\prod_{j=1}^{n_{\Ai+\Ai}}\gcd(N,s_j)\times N^{n_{\Bi}}$. Substituting it into~\eqref{42} yields
\begin{equation}
    \eqref{42}=\frac{N^{n_{\Ai}}}{\prod_{i=1}^{n_{A_I}}\gcd(N,r_i)}\;(\vcentcolon=\Gamma_4).
    \end{equation}
\par
Overall, 
\begin{eqnarray}
    \Gamma_A=\Gamma_1\times\Gamma_3\nonumber\\
    \Gamma_B=\Gamma_2\times\Gamma_3\times\Gamma_4.
\end{eqnarray}
Referring to~\eqref{eqn:EEequalsuperposition} and ~\eqref{nini}, we finally arrives at
\begin{eqnarray}
S_A&=&\biggl[\log\left(\frac{|G|}{|\tilde{G}_A||\tilde{G}_B|}\right)-2l_y\log\left(\prod_{i=1}^{n_{\Ai}}\gcd\bigl(N,r_i)\right)-l_y(n_{B_I}-n_{A_I})\log N\biggr]\nonumber\\
    &-&\log\biggl(\frac{N^{n_{A_I}}}{\prod_{i=1}^{n_{\Ai}}\gcd\bigl(N,r_i)}\biggr)-2\log\left(\prod_{j=1}^{n_{\Ai}+n_{\Ao}}\gcd(N,s_j)\right)\;\;(n_{\Ai}\leq n_{\Bi}).
\end{eqnarray}
Proof of the theorem~\eqref{main2} is completed by rewriting $\frac{|G|}{|\tilde{G}_A||\tilde{G}_B|}=N^{(Area)}$, where $(Area)$ denote the number of vertex operators that have support both on $A$ and $B$, which is interpreted as the number of vertices surrounding the subsystem $A$. 
Analogous argument leads to~\eqref{main2} in the case of $n_{\Bi}<n_{\Ai}$. \hfill $\square$\par
The first three terms depend on the system size, whereas
the sub-leading order terms corresponding to the last two terms, depend on the number of clock states inside A which are connected with B (i.e., $n_{A_I}$) and invariant factors of the Laplacian of the sub-graph.

While we corroborate that the sub-leading order term of the entanglement entropy~\eqref{main} in the case of the square lattice is topological by an intuitive argument given in Sec.~\ref{sec4}, one can understand the topological origin of the sub-leading term in the case of the square lattice from a different perspective based on the result~\eqref{main2}. 
In the case of the square lattice, for a disk geometry, we have $n_{A_I}=n_{B_I}=2$ and invariant factors of the sub-graph (i.e., disk geometry) become trivial (namely $r_i=s_j=1$). Thus,~\eqref{main2} is identical to~\eqref{main}. The sub-leading term, $-n_{A_I}\log N=-2\log N$ retains the same value, regardless of the width of the disk as the number $n_{A_I}$ is always constant $2$.

\bibliographystyle{ieeetr}
\section{Brief comments on other cases of the higher rank topological phases}\label{sec5}
We have studied entanglement entropy of the higher rank topological phases defined in~\eqref{hamiltonian} by making use of formalism of the stabilizers~\cite{Hamma2005} jointly with the one of the Laplacian. One could study entanglement entropy of other higher rank topological phases by resorting to the similar approach.

One example is the model studied recently in~\cite{PhysRevB.107.125154}, which is introduced as follows. At each vertex of the square lattice, we introduce two clock states $\ket{a}_{(x,y)}\ket{b}_{(x,y)}$, $a,b\in\mathbb{Z}_N$. Also, 
 \begin{figure}[h]
    \begin{center}
         \includegraphics[width=0.35\textwidth]{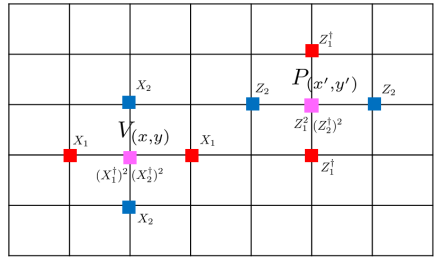}
       \end{center}
       \caption{Two types of terms defined in~\eqref{VP}.
 }
        \label{plane}
   \end{figure}
define operators acting on these states as (recall that $\omega=e^{2\pi i/N}$)
\begin{eqnarray}
Z_{1,(x,y)}\ket{a}_{(x,y)}\ket{b}_{(x,y)}=\omega^a\ket{a}_{(x,y)}\ket{b}_{(x,y)},\;Z_{2,(x,y)}\ket{a}_{(x,y)}\ket{b}_{(x,y)}=\omega^b\ket{a}_{(x,y)}\ket{b}_{(x,y)}\nonumber\\
X_{1,(x,y)}\ket{a}_{(x,y)}\ket{b}_{(x,y)}=\ket{a+1}_{(x,y)}\ket{b}_{(x,y)},\;X_{2,(x,y)}\ket{a}_{(x,y)}\ket{b}_{(x,y)}=\ket{a}_{(x,y)}\ket{b+1}_{(x,y)}.
\end{eqnarray}
Introducing the following terms by (Fig.~\ref{plane})
\begin{eqnarray}
V_{(x,y)}\vcentcolon=X_{1,(x+1,y)}X_{1,(x-1,y)}(X_{1,(x,y)}^{\dagger})^2X_{2,(x,y+1)}X_{2,(x,y-1)}(X_{2,(x,y)}^{\dagger})^2\nonumber\\
P_{(x,y)}\vcentcolon=Z_{1,(x,y+1)}^{\dagger}Z_{1,(x,y-1)}^{\dagger}Z^2_{1,(x,y)}Z_{2,(x+1,y)}Z_{2,(x-1,y)}(Z_{2,(x,y)}^{\dagger})^2,\label{VP}
\end{eqnarray}
Hamiltonian reads
\begin{equation}
    H_{\mathbb{Z}_N}=-\sum_{x,y}(V_{(x,y)}+P_{(x,y)})+h.c\label{zn}.
\end{equation}
The GSD of the phase on the torus geometry with length of the lattice in the $x$ and $y$ direction being $n_x$ and $n_y$, is found to be~\cite{PhysRevB.107.125154}
\begin{equation}
    GSD=[N\times\gcd(N,n_x)\times\gcd(N,n_y)\times\gcd(N,n_x,n_y)]^2.
\end{equation}
By making use of the similar line of thoughts outlined in Sec.~\ref{sec3}, one obtains the entanglement entropy of the disk geometry as
\begin{equation}
    \boxed{S_A=(Area)\log N-4\log N.}
\end{equation}
Here, $Area$ is the number of vertices surrounding the disk. 
Compared with the result~\eqref{main}, the absolute value of the topological entanglement entropy is increased. Also, the total quantum dimension of the model is given by $\sqrt{\sum_ad^2_a}=[N\times\gcd(N,n_x)\times\gcd(N,n_y)\times\gcd(N,n_x,n_y)]$, thus, analogously to~\eqref{main}, the relation between total quantum dimension and the topological entanglement entropy does not hold.

It would be interesting to study other higher rank topological phases, such as the ones studied in~\cite{PhysRevB.97.235112,PhysRevB.106.045145,half2022,gorantla2022gapped,PhysRevB.107.155151}. Generically, we make a conjecture
that one cannot associate the topological entanglement entropy in the higher rank topological phases with the total quantum dimension of the fractional excitations and also that 
the absolute value of the topological entanglement entropy becomes larger when the model contains the terms which involves clock states in the longer range.
We leave confirmation of this speculation for
future studies.

\section{Conclusion}\label{sec6}
In this work, we study entanglement entropy of unusual topological stabilizer models 
with dipole symmetry, which is one of the multipole symmetries, admitting dipole of the fractional charges. As opposed to the entanglement entropy of the fracton topological phases, where it is cumbersome to identify the entanglement entropy due to the sub-system symmetries~\cite{PhysRevB.97.125102,PhysRevB.97.134426,nonlocal2018}, entanglement entropy of our model is easily obtained based on the formulation of the combinatrics.
While the GSD and the total quantum dimension of the fractional excitations drastically changes depending on the system size, topological entanglement entropy $\gamma$ takes a constant, given by $\gamma=-2\log N$. Due to this result, the well-known relation between the topological entanglement entropy and the total quantum dimension is not valid in the case of the higher rank topological phases. 
Such a result can be understood by decomposition of the model into two or rearrangement of the lattice in the simplest case $N=2$. We further study the entanglement entropy of the model on generic lattices and have found that the entanglement entropy depends on the number of clock states surrounding the system and invariant factors of the Laplacian of the sub-graph. The result presented in this work complies with the growing interests in topological phases with multipole symmetries in view of quantum entanglement and combinatrics. 
\par
There are several future research directions regarding the present study. 
It is known that capability of the error correction of the toric code is characterized by the classical Ising universality class~\cite{dennis2002topological}. 
It would be interesting to study how our model~\eqref{hamiltonian} can be utilized for quantum error corrections, in particular, to identify what kind of universality class characterizes the capability of the error correction in our model. One naively expect that due to the fact that the model has the multipole symmetries, such a universal class is qualitatively different from the one in the conventional toric code.
While we focus on the entanglement entropy in the higher rank topological phases, one would be curious to elucidate behavior of other observable to quantify the quantum entanglement. One candidate would be entanglement negativity~\cite{eisert1999comparison,Vidal2002}, an important observable to extract quantum correlation rather than classical one. It would be intriguing to investigate the entanglement negativity of our model to see how different it is compared with the case of the conventional topologically ordered phases~\cite{Vidal_negativity}.

\section*{Acknowledgement}
The author thanks Bo Han, Ken Shiozaki, Taiichi Nakanishi, Masazumi Honda, Yuval Oreg for helpful discussion. Author also thanks Guilherme Delfino and Jung Hoon Han for notification of their papers~\cite{half2022,PhysRevB.107.155151}.
This work is in part supported byKAKENHI-PROJECT-23H01097.


\bibliography{main}
\appendix
\section{Proof of Eq.~(\ref{eqn:EE})}
\label{appendix:EEproof}
Given $g,g' \in G$, we have $g'_A = g_A \Leftrightarrow g' = h g, h \in G_B$. 
Then~(\ref{eqn:reduced2}) is reduced to
\begin{align}
\rho_A &= \frac{|G_B|}{|G|} \sum_{g \in G/G_B, g' \in G_A} g_A |0\rangle_A {}_A \langle 0 | (g_A g'_A)^\dagger.
\end{align}
We have
\begin{align}
\rho^2_A &= \biggl(\frac{|G_B|}{|G|}\biggr)^2 \sum_{g,\tilde{g} \in G/G_B, g',\tilde{g}' \in G_A} g_A |0\rangle_A {}_A \langle 0| (g_A \tilde{g}_A)^\dagger  g'_A |0\rangle_A {}_A \langle 0| (g'_A \tilde{g}'_A)^\dagger \nonumber \\
&= \biggl(\frac{|G_B|}{|G|}\biggr)^2 \sum_{g \in G/G_B, \tilde{g},\tilde{g}' \in G_A} g_A |0\rangle_A {}_A \langle 0|(g_A \tilde{g}_A \tilde{g}'_A)^\dagger \nonumber \\
&= \biggl(\frac{|G_B|}{|G|}\biggr)^2 |G_A| \sum_{g \in G/G_B, \tilde{g} \in G_A} g_A |0\rangle_A {}_A \langle 0| (g_A \tilde{g}_A)^\dagger  \nonumber \\
&= \frac{|G_B|}{|G|} |G_A| \rho_A,
\end{align}
from which one finds
\begin{align}
\rho^n_A &= \left( \frac{|G_A|\cdot |G_B|}{|G|} \right)^{(n-1)} \rho_A.
\end{align}
The entanglement entropy is obtained as
\begin{align}
S_A &= \lim_{n \to 1} \frac{1}{1-n} \log{\text{tr} \rho^n_A} = \log \frac{|G|}{|G_A|\cdot |G_B|}.
\end{align}

\section{Entanglement entropy with respect to the generic ground state}
\label{appendix:higherrankee}

In this section, we give a derivation of the entanglement entropy of three subsystems discussed in Sec.~\ref{singlerowco} in the case of the generic ground state~\eqref{generic}.
The density matrix $\rho_A^{\zeta}$ of a subsystem $A$ in our model with respect to the generic ground state~\eqref{generic} reads
\begin{equation}
  \rho_A^{\zeta}= \text{Tr}_B\ket{\zeta}\bra{\zeta},
\end{equation}
where $\ket{\zeta}$ is given by~\eqref{generic}. From~\eqref{gs}, it is explicitly written as
\begin{equation}
    \rho_A^{\zeta}=\sum_{\substack{a,b,c,d\\ a^\prime,b^\prime,c^\prime,d^\prime}}\alpha_{ab,cd}\bar{\alpha}_{a^\prime b^\prime,c^\prime d^\prime}\text{Tr}_B\bigl[(\eta^x_{1})^a(\gamma^x_{1})^b(\eta^x_{2})^c(\gamma^x_{2})^d\rho_0(\eta^x_{1})^{a^\prime}(\gamma^x_{1})^{b^\prime}(\eta^x_{2})^{c^\prime}(\gamma^x_{2})^{d^\prime}\bigl]\label{zetaa}
\end{equation}
with $\rho_0 =|\zeta_{00}\rangle \langle \zeta_{00} |$.
Below we look at the three cases corresponding to Fig.~\ref{fig:singlerow}a,b,c. 
\subsection{Single row I}
In Fig.~(\ref{fig:singlerow})(a), recalling the form of the logical operators~\eqref{logi} (see also Fig.~\ref{logical}), 
the logical operator $\eta^x_{1}$ and $\g^x_{1}$ act within subsystem B, hence the term $\text{Tr}_B[\cdots]$ in~\eqref{zetaa} is transformed as
\begin{equation}
  \frac{1}{|G|} \sum_{g,g'\in G} (\eta^x_{2A})^c(\gamma^x_{2A})^dg_A |0\rangle_A {}_A \langle 0| (g_A g'_A)^\dagger (\eta^{x\dagger}_{2A})^{c^\prime}(\gamma^{x\dagger}_{2A})^{d^\prime} \cdot {}_B \langle 0 |g'^\dagger_B (\eta^x_1)^{a-a^\prime} (\gamma^x_1)^{b-b^\prime} (\eta^{x}_{2B})^{c-c^\prime} (\gamma^{x}_{2B})^{d-d^\prime} |0\rangle_B\label{hid}
\end{equation}
where rewrite the logical operator by the product form $\eta_2^x=\eta_{2A}^x\otimes\eta_{2B}^x$, and similarly for $\gamma_{2}^x$. From the inner product of the state $\ket{0}_B$ in~\eqref{hid}, one finds that 
\begin{equation}
    g^\prime_B=I_B,\;a=a^\prime,b=b^\prime,c=c^\prime,d=d^\prime.
\end{equation}
Hence, the density matrix $\rho_A^{\zeta}$ becomes
\begin{equation}
    \rho_A^{\zeta}=\sum_{c,d}\sum_{g\in G,g^\prime\in G_A}\frac{1}{|G|}\sum_{a,b}|\alpha_{ab,cd}|^2(\eta_2^x)^c(\gamma_2^x)^dg_A |0\rangle_A {}_A \langle 0|(\eta_2^{x\dagger})^c(\gamma_2^{x\dagger})^d (g_A g'_A)^\dagger.
\end{equation}
From this form we obtain
\begin{equation}
    \text{Tr}_A[ (\rho_A^{\zeta})^n]=\sum_{c,d}\biggl[\sum_{a,b}|\alpha_{ab,cd}|^2\biggr]^n\times\biggl(\frac{|G_A||G_B|}{|G|}\biggr)^{n-1}. 
\end{equation}
The entanglement entropy is given by
\begin{equation}
    S_A=\lim_{n \to 1} \frac{1}{1-n} \log{\text{tr}_A (\rho_A^{\zeta})^n}=\log\frac{|G|}{|G_A||G_B|}-\sum_{c,d}\biggl[\biggl(\sum_{a,b}|\alpha_{ab,cd}|^2\biggl)\log\biggl(\sum_{a,b}|\alpha_{ab,cd}|^2\biggl) \biggr].
\end{equation}
Therefore, \eqref{r1prime} follows.
\subsection{Single row II}
In the case of Fig.~(\ref{fig:singlerow})(b), from~\eqref{logi} (see also Fig.~\ref{logical}), it follows that two logical operators running in the $y$-direction, $\eta^x_2$ and $\gamma_2^x$ exclusively act on B. Regarding other two, $\eta^x_1$ and $\gamma_1^x$ which go around in the $x$-direction, 
one can deform these operators so that they act within A. Therefore the term $\text{Tr}_B[\cdots]$ in~\eqref{zetaa} becomes
\begin{equation}
     \frac{1}{|G|} \sum_{g,g'\in G} (\eta^x_{1})^a(\gamma^x_{1})^bg_A |0\rangle_A {}_A \langle 0| (g_A g'_A)^\dagger (\eta^{x\dagger}_{1})^{a^\prime}(\gamma^{x\dagger}_{1})^{b^\prime} \cdot {}_B \langle 0 |g'^\dagger_B  (\eta^{x}_{2})^{c-c^\prime} (\gamma^{x}_{2})^{d-d^\prime} |0\rangle_B,
\end{equation}
from which one finds
\begin{equation}
    g'_B=I_B,\;c=c^\prime,d=d^\prime.
\end{equation}
The density matrix is then rewritten as
\begin{equation}
     \rho_A^{\zeta}=\sum_{a,b,a^\prime,b^\prime}\sum_{c,d}\sum_{g\in G,g^\prime\in G_A}\frac{\alpha_{ab,cd}\bar{\alpha}_{a^\prime b^\prime,cd}}{|G|}(\eta_1^x)^a(\gamma_1^x)^bg_A |0\rangle_A {}_A \langle 0|(\eta_1^{x\dagger})^{a^\prime}(\gamma_1^{x\dagger})^{b^\prime} (g_A g'_A)^\dagger\label{zetab}
\end{equation}
Introducing orthogonal states by 
\begin{equation}
    \ket{p,q}_A\vcentcolon=\frac{1}{\sqrt{N\times\gcd(N,n_x)}}\sum_{a=0}^{\gcd(n,n_x)-1}\sum_{b=0}^{N-1}\nu^{pa}\omega^{qb}(\eta_1^x)^a(\gamma_1^x)^b |0\rangle_A,\;(0\leq p\leq\gcd(N,n_x)-1,0\leq q\leq N-1)
\end{equation}
with $\nu=e^{2\pi i/\gcd(N,n_x)}$, $\omega=e^{2\pi i/N}$, \eqref{zetab} is transformed as
\begin{equation}
      \rho_A^{\zeta}=\frac{1}{|G|}\sum_{g\in G,g^\prime\in G_A}\sum_{p=0}^{\gcd(N,n_x)-1}\sum_{q=0}^{N-1}\biggl(\frac{1}{\Gamma}\sigma_{kl}\nu^{kp}\omega^{lq}\biggr)g_A |p,q\rangle_A {}_A \langle p,q|(g_A g'_A)^\dagger,
\end{equation}
where 
\begin{equation}
    \sigma_{kl}\vcentcolon= \sum_{\substack{a,a^\prime,b,b^\prime,c,d \\ a-a^\prime=k\mod \gcd(N,n_x)\\b-b^\prime=l\mod N}}\alpha_{ab,cd}\bar{\alpha}_{a^\prime b^\prime,cd}.
\end{equation}
Recall that we have defined $\Gamma=N\times\gcd(N,n_x)$.
Defining $\lambda_{p,q}\vcentcolon=\frac{1}{\Gamma}\sigma_{kl}\nu^{kp}\omega^{lq}$, we have
\begin{equation}
     \text{Tr}_A[ (\rho_A^{\zeta})^n]=\sum_{p,q}\lambda_{p,q}^n\times\biggl(\frac{|G_A||G_B|}{|G|}\biggr)^{n-1}.
\end{equation}
Substituting this into
\begin{equation}
  S_A= \lim_{n \to 1} \frac{1}{1-n} \log{\text{tr}_A (\rho_A^{\zeta})^n} \label{trd}
\end{equation}
yields~\eqref{nu1}.
\subsection{Single column}
In the case of Fig.~(\ref{fig:singlerow})(c), 
one can deform the logical operators~\eqref{logi} so that $\eta^x_1$ and $\eta_2^x$ act exclusively on B and $\gamma_2^x$ acts within A, with $\gamma_1^x$ acting on both of A and B. The term $\text{Tr}_B[\cdots]$ in~\eqref{zetaa} is described by
\begin{equation}
    \frac{1}{|G|} \sum_{g,g'\in G} (\gamma^x_{1A})^b(\gamma^x_{2})^dg_A |0\rangle_A {}_A \langle 0| (g_A g'_A)^\dagger (\gamma^{x\dagger}_{1A})^{b^\prime}(\gamma^{x\dagger}_{2})^{d^\prime} \cdot {}_B \langle 0 |g'^\dagger_B  (\eta^{x}_{1})^{a-a^\prime} (\gamma^{x}_{1B})^{b-b^\prime}(\eta^{x}_{2})^{c-c^\prime} |0\rangle_B\label{b15}
\end{equation}
The inner product gives a constraint
\begin{equation}
    g'_B=I_B,\;a=a^\prime, b=b^\prime,c=c^\prime.
\end{equation}
Define orthogonal states by
\begin{equation}
    \ket{b,q}_A\vcentcolon=\frac{1}{\sqrt{N}}\sum_{d=0}^{N-1}\omega^{qd}(\gamma^x_{1A})^b(\gamma^x_{2})^d |0\rangle_A,
\end{equation}
the density matrix $\rho^{\zeta}_A$ is rewritten as
\begin{equation}
    \rho^{\zeta}_A=\frac{1}{|G|}\sum_{g\in G,g^\prime\in G_A}\sum_{b,q}\lambda^{(b)}_qg_A |b,q\rangle_A {}_A \langle b,q|(g_A g'_A)^\dagger
\end{equation}
with 
\begin{equation}
    \lambda_{q}^{(b)}=\frac{1}{N}\sum_{k=0}^{N-1}\omega^{kq}\sigma^{(b)}_k,\;\;\sigma^{(b)}_k=\sum_{\substack{a,c,d,d^\prime\\ d-d^\prime=k\mod N}}\alpha_{ab,cd}\bar{\alpha}_{ab,cd^\prime}.\label{nu44}
\end{equation}
By calculating~\eqref{trd}, one obtains~\eqref{nu3}.
\end{document}